# Ψ - MODEL OF MICRO- AND MACROSYSTEMS

**E.E. Perepelkin[a], B.I. Sadovnikov[a], N.G. Inozemtseva[b]**

[a] *Faculty of Physics, Lomonosov Moscow State University, Moscow, 119991 Russia*
*E. Perepelkin e-mail: pevgeny@jinr.ru, B. Sadovnikov sadovnikov@phys.msu.ru*
[b] *Dubna State University, Moscow region, Moscow,141980 Russia*
*e-mail: nginozv@mail.ru*

**Abstract**

A mathematical model (referred as Ψ - model for convenience) has been developed, which allows describing certain class of micro- and macrosystems. Ψ - model is based on continuum and quantum mechanics. Ψ - model describes micro- and macrosystems, in which vector field of velocities of probability flows, charge, mass has specific spiral structure. The field of velocities has spiral structure on concentric spherical surfaces. The velocity field is not defined and has a characteristic property on the poles of sphere and on the axis and tends to zero at infinity. The behavior of Ψ - model can be described in the general case with unsteady periodic singular solution of the Schrödinger equation. The application of Ψ - model to the description of elementary particles gives naturally the intrinsic magnetic moment (spin) without need in modification of $g$ -factor. The spin is interpreted as rotation in the framework of the relativity theory. The application of Ψ -model to astrophysics gives formally an interpretation of black and white hole, wormhole, structure of the galaxy, anti-galaxy and «enclosed worlds».

**Introduction**

The Schrödinger equation is one of the basic equations of quantum mechanics [1]. There are different methods of solving the Schrödinger equation: the method of perturbation theory [2-6], variational method [7], the Monte-Carlo method [8-12], the method of density functional theory [13-17], the WKB approximation and semi-classical expansion [18-20], the Hartree-Fock method [21-22]. Besides the listed, there are exact analytical solutions for some quantum mechanical systems [23-26].

In most cases, there is a need to solve the Schrödinger equation for the given physical system and the potential related to it with numerical methods. The existence of exact solutions of the Schrödinger equation is quite useful for the process of education and obtaining the intuitive perceptions of quantum system structure.

Exact singular solutions of the three-dimensional unsteady Schrödinger equation have been found in the paper with, in general, various kinds of spherically asymmetric potentials. The method of searching exact solution is based on the method of continuum mechanics, which we described in the paper [27]. The principle of the method consists in the direct relationship between the Schrödinger equation for the wave function Ψ

$$\frac{i}{\beta}\frac{\partial \Psi}{\partial t} = -\alpha\beta\left(\hat{p} - \frac{\gamma}{2\alpha\beta}\vec{A}\right)^2 \Psi + \frac{1}{2\alpha\beta}\frac{\left|\gamma\vec{A}\right|^2}{2}\Psi + U\Psi, \qquad (i.1)$$

where

$$U(\vec{r},t) = -\frac{1}{\beta}\left\{\frac{\partial \varphi(\vec{r},t)}{\partial t} + \alpha\left[\frac{\Delta\sqrt{f(\vec{r},t)}}{\sqrt{f(\vec{r},t)}} - \left|\nabla\varphi(\vec{r},t)\right|^2\right] + \gamma\left(\vec{A},\nabla\varphi\right)\right\}, \qquad (i.2)$$

and continuity equation for probability density $f = |\Psi|^2$:

$$\frac{\partial f}{\partial t} + \operatorname{div}\left[f \langle \vec{v} \rangle\right] = 0 \quad \text{or} \quad \frac{dS}{dt} = Q, \qquad (i.3)$$

where

$$\langle \vec{v} \rangle = -\alpha \nabla \Phi + \gamma \vec{A}, \qquad (i.4)$$

$$i\Phi(\vec{r},t) = \operatorname{Ln}\left[\frac{\Psi}{\bar{\Psi}}\right] = \ln\left|\frac{\Psi}{\bar{\Psi}}\right| + i\operatorname{Arg}\left[\frac{\Psi}{\bar{\Psi}}\right] = i2(\varphi(\vec{r},t) + \pi k), \qquad (i.5)$$

$$S = -\ln f, \quad Q = \operatorname{div}\langle \vec{v} \rangle.$$

As equation (i.3) has a wide field of application in classic mechanics, hydro- and aerodynamics, statistical mechanics, quantum mechanics, the values of constants $\alpha, \beta, \gamma$ may depend on the physics field of consideration. For the quantum system case, we will use hereinafter the designations from the paper [27]: $\alpha = -\frac{\hbar}{2m}$, $\beta = \frac{1}{\hbar}$, $\gamma = -\frac{e}{m} = -\frac{q_e}{m_e}$.

Value $S = -\ln f$ is associated with continuous entropy $H$

$$H = -\int_{(\infty)} f \ln f \, d^3r = -\int_{(\infty)} f S \, d^3r = \langle S \rangle. \qquad (i.6)$$

The following equation has been obtained in this paper for entropy $H$ due to (i.3)

$$\frac{dH}{dt} = \langle Q \rangle. \qquad (i.7)$$

Unlike classical approach to solving initial boundary value problem for equation (i.1), when potential (i.2), initial and boundary conditions are specified and it is required to find wave function $\Psi$, initial or boundary conditions for probability density function $f = |\Psi|^2$ and vector field of probability flow $\langle \vec{v} \rangle$ (i.4) are reference conditions here.

For this paper, the field $\langle \vec{v} \rangle$ is chosen as a gradient of phase $\varphi$ of wave function $\Psi$. Such an expression in the spherical coordinate system $r, \theta, \phi$ is as follows:

$$\varphi = k\phi + n\theta + c_0, \quad n, k \in \mathbb{Z} \quad (or \ n, k \in \mathbb{R}) \qquad (i.8)$$

$$\langle \vec{v} \rangle = -2\alpha \nabla \varphi = -2\alpha \left(\frac{k}{r \sin \theta} \vec{e}_\phi + \frac{n}{r} \vec{e}_\theta\right) \qquad (i.9)$$

The mentioned expression for probability flow velocity $\langle \vec{v} \rangle$ has solenoidal $\frac{k}{r \sin \theta} \vec{e}_\phi$ and irrotational $\frac{n}{r} \vec{e}_\theta$ component of the velocity and is, in fact, «scalar» (i.8) analogue of the Helmholtz decomposition (i.9). The solenoidal component of the expression (i.9) is consistent with the principle of Bohr–Sommerfeld quantization and produces magnetic field of the following form

$$\vec{B} = -\frac{q_m^{(Wb)} \delta(\rho)}{2\pi\rho} \vec{e}_z, \qquad (i.10)$$

where

$$q_m^{(Wb)} q_e = 2\pi\hbar k, \quad \rho = r\sin\theta.$$

The value $q_m^{(Wb)}$ coincides with the known expression for the magnet charge and satisfies the quantization condition [28]. Notice that the magnetic field associated to the «magnetic charge» $q_m^{(Wb)}$ satisfies classical Maxwell's equations, that is $\operatorname{div}\vec{B} = 0$. Intrinsic magnetic moment of such quantum system is of the form

$$\vec{\mu}_s = \mu_B k \vec{e}_z, \qquad (i.11)$$

and at $k = \pm 1$ coincides with the intrinsic magnetic moment of the electron with the spin $s = \pm\frac{1}{2}$.

Thus, the choice of the probability flow density $\vec{J} = ef\langle\vec{v}\rangle$ (i.9) has a number of aspects of consideration interesting for theoretical physics. For example, the paper describes the mathematical model of an electron-like elementary particle. Such quantum system has the same as the electron electric charge $q_e$, mass $m_e$, intrinsic magnetic moment (spin).

From geometry point of view, expression (i.3) determines the flow along spiral trajectories from one pole of the sphere to the opposite one. The mentioned spiral trajectories are characteristics (motion equations)

$$\xi(\theta,\phi) = \phi + \frac{k}{n}\operatorname{ctg}\theta, \qquad (i.12)$$

for a hyperbolic-type equation corresponds to the Schrödinger equation (i.1). As a result, the exact solution of the Schrödinger equation is solved by the characteristics method and depends on the initial distribution $f_0(r,\phi,\theta)$ of probability density

$$\Psi(\vec{r},t) = \frac{1}{\sqrt{\sin\theta}} F_0\left(r, \phi + \frac{k}{n}\left(\operatorname{ctg}\theta - \operatorname{ctg}\left(\theta + \frac{2\alpha n}{r^2}t\right)\right), \theta + \frac{2\alpha n}{r^2}t\right) e^{i\left(n\theta - \frac{E}{\hbar}t\right)}, \qquad (i.13)$$

$$f(r,\phi,\theta,t)\big|_{t=0} = \frac{F_0(r,\phi,\theta)}{\sin\theta} = f_0(r,\phi,\theta), \qquad (i.14)$$

One can see from the form of solution (i.13) that it is periodic due to the periodicity of function $ctg$. The frequency and period of such a process depend on the sphere radius and are of the following form:

$$\omega(r) = \frac{2|\alpha|n}{r^2} = \frac{\hbar n}{mr^2}, \quad T(r) = \frac{m}{\hbar n}\pi r^2. \qquad (i.15)$$

The paper shows that potential (i.2) may contain potential wells separated from each other by potential barriers. As a result, the highest probability will be focused in the region of potential (i.2)'s wells.

The first part of the paper reports the main theoretical results.

We have conducted in paragraph 1.1 the research of properties of the vector field of probability flow (i.4), have obtained expression for the magnetic field (i.10) and intrinsic magnetic moment (i.11) and have found an exact solution (i.13), (i.14) of the unsteady Schrödinger equation (i.1), its periodicity has been considered as well (i.15).

A mathematical model of an elementary particle of the form of a toroidal ring has been developed in paragraph 1.2. The probability density function $f$ has a distribution structure of a toroidal ring form. At that, the velocity of probability flow (i.9) has only one azimuthal angle ($n = 0$). The developed model possesses charge, mass and intrinsic magnetic moment (i.9) same as the electron's one, besides the rotation velocity $\langle v \rangle < c$, i.e. is compatible to the relativity theory and does not require modification of $g$-factor for the transition from classical to quantum mechanics.

Equation (i.7) has been obtained in paragraph 1.3 for continuous entropy $H = \langle S \rangle$ (i.6), its regions of increase, decrease and constancy have been studied.

The second part considers examples of micro- and macrosystems based on the theoretical results from the first part.

A quantum system with modified Coulomb potential has been considered in paragraph 2.1 as a passage to the limit between quantum mechanics and continuum mechanics. A system with periodic boundary conditions for the probability density function has been considered in paragraph 2.2; the potential has spiral-shaped wells separated by spiral potential barriers and the maximum probability is focused in the regions of potential wells.

The transfer has been considered in paragraph 2.3 of the developed mathematical model ($\Psi$-model) to the astrophysics. Spiral trajectories (i.12) are formally considered as arms of the galaxy entering (leaving) poles of the sphere, which are formally the «black hole» and «white hole» respectfully. The behavior of entropies $S$ and $H$ has been considered in the neighborhood of the «black» and «white holes». The interpretation has been shown formally of the «enclosed worlds» system as concentric spheres connected by one axis (wormhole) and having their own significant period (i.15).

## §1 Theoretical results

### 1.1 An exact solution

Let us consider particular solutions for equation of continuity (1.1.1), and as consequence for the Schrödinger or Pauli equations [27].

$$\frac{\partial f}{\partial t} + \operatorname{div}\left[f \langle \vec{v} \rangle\right] = 0. \tag{1.1.1}$$

$$\frac{dS}{dt} = S_t + \left(\langle \vec{v} \rangle, \nabla S\right) = \operatorname{div}\langle \vec{v} \rangle = Q, \tag{1.1.2}$$

where $S \stackrel{\text{det}}{=} -\ln f$. We express velocity of probability flow $\langle \vec{v} \rangle$ according to paper [27] in the following form (Helmholtz decomposition)

$$\langle \vec{v} \rangle = -\alpha \nabla \Phi + \gamma \vec{A}, \tag{1.1.3}$$

where $\vec{A}$ corresponds to the solenoidal component of the velocity field.

Paper [32] showed that field $\nabla\Phi$ can be solenoidal as well for non-smooth scalar potential $\Phi$, for example

$$\Phi = 2\varphi = c_1\phi + c_2, \qquad (1.1.4)$$

$$\nabla\Phi = \frac{c_1}{\rho}\vec{e}_\phi = \frac{c_1}{r\sin\theta}\vec{e}_\phi,$$

$$\mathrm{rot}[\nabla\Phi] \neq \vec{0},$$

where $\phi$ is azimuthal angle, $\rho = r\sin\theta$ is azimuthal radius, and $c_1, c_2$ are constant values. Notice that field $\mathrm{rot}[\nabla\Phi]$ is different from zero only on the OZ axis [32].

Let us consider extended for of potential (1.1.4) for three-dimensional case

$$\varphi(\theta,\phi) = n\theta + k\phi + c_0, \qquad (1.1.5)$$

where $k, n, c_0$ are constant values. In special case, at $n=0$, expression (1.1.5) transforms into (1.1.4). The velocity of probability flow (1.1.3) has the following form in the spherical coordinate system taking into account (1.1.5):

$$\langle\vec{v}\rangle = -\alpha\nabla\Phi = -2\alpha\left(\frac{k}{r\sin\theta}\vec{e}_\phi + \frac{n}{r}\vec{e}_\theta\right) \qquad (1.1.6)$$

Having calculated the divergence of velocity field $Q$ (1.1.2), we obtain

$$Q = -2\alpha\left(\frac{1}{r^2\sin\theta}\frac{\partial}{\partial\theta}\sin\theta\frac{\partial}{\partial\theta}(n\theta) + \frac{1}{r^2\sin^2\theta}\frac{\partial^2}{\partial\phi^2}(k\phi)\right),$$

$$Q = -\frac{2\alpha n}{r^2}\mathrm{ctg}\,\theta. \qquad (1.1.7)$$

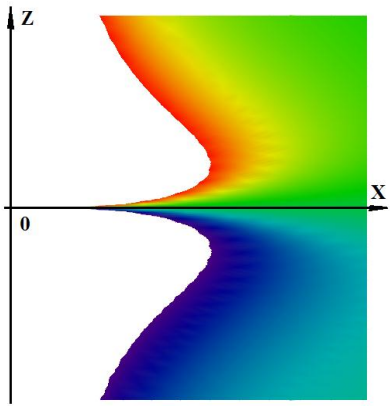

Fig.1 Distribution of Q

Value $Q$ corresponds to the density of the source of field $\langle\vec{v}\rangle$ from the physical standpoint, and as $Q$ is different from zero in general case (1.1.7), there are such sources. Expression (1.1.7) becomes zero at $\theta = \frac{\pi}{2}$ and tends to zero at infinity ($r\to\infty$). Note that density (1.1.7) is symmetric in respect to azimuthal angle $\phi$.

Fig. 1 shows the distribution of source density $Q$ in the XOZ plane, that is at $\phi = 0$. It is clear from Fig. 1 that when going over the angle $\theta = \frac{\pi}{2}$ field sources reverse their signs to the opposite.

Thus, the vector field $\langle\vec{v}\rangle$ corresponds to the flow from one part of the semispace $Z < 0$ to the other part of the semispace $Z > 0$. The sign of the constants in expression (1.1.7) determines the flow direction.

Let us show that the field $\langle \vec{v} \rangle$ is a solenoidal one. Having calculated the curl (rot) from expression (1.1.6), we obtain

$$\text{rot}\langle \vec{v} \rangle = \frac{-2\alpha}{r\sin\theta}\left[\frac{\partial}{\partial\theta}\left(\frac{k}{r\sin\theta}\sin\theta\right) - \frac{\partial}{\partial\phi}\frac{n}{r}\right]\vec{e}_r + \frac{-2\alpha}{r}\left[-\frac{\partial}{\partial r}\left(r\frac{k}{r\sin\theta}\right)\right]\vec{e}_\theta +$$

$$+ \frac{-2\alpha}{r}\left[\frac{\partial}{\partial r}\left(r\frac{n}{r}\right)\right]\vec{e}_\phi = \frac{-2\alpha}{r\sin\theta}\left[\frac{\partial}{\partial\theta}\left(\frac{k}{r}\right) - \frac{\partial}{\partial\phi}\frac{n}{r}\right]\vec{e}_r + \frac{2\alpha}{r}\frac{\partial}{\partial r}\left(\frac{k}{\sin\theta}\right)\vec{e}_\theta - \frac{2\alpha}{r}\frac{\partial n}{\partial r}\vec{e}_\phi = \vec{0}. \quad (1.1.8)$$

From (1.1.8) it follows that $\text{rot}\langle \vec{v} \rangle = \vec{0}$ is zero almost everywhere except the null set (negligible set). In order to show that on the OZ axis $\text{rot}\langle \vec{v} \rangle$ is different from zero, we calculate $\oint_\gamma \langle \vec{v} \rangle d\vec{l}$ the circulation of vector $\langle \vec{v} \rangle$ around closed curve $\gamma$ in the shape of circle belonging to the sphere with radius $r$ (see Fig. 2)

$$d\vec{l} = \vec{e}_\theta r d\theta + \vec{e}_\phi r\sin\theta d\phi,$$

$$\oint_\gamma \langle \vec{v} \rangle d\vec{l} = -2\alpha k \int_{\phi_0}^{\phi_0+2\pi}\frac{r\sin\theta}{r\sin\theta}d\phi - 2\alpha n\left(\int_{\theta_0}^{\pi-\theta_0}\frac{r}{r}d\theta + \int_{\pi-\theta_0}^{\theta_0}\frac{r}{r}d\theta\right) = -4\pi\alpha k, \quad (1.1.9)$$

which yield according to the Stokes' theorem

$$\int_{\Sigma_\gamma}\text{rot}\langle \vec{v} \rangle d\vec{\sigma} = -4\pi\alpha k. \quad (1.1.10)$$

The right part of expression (1.1.10) does not contain information on the value of area $\Sigma_\gamma$. Thus, path $\gamma$ may be shrunk as much as desired around the OZ axis, at that, integral value (1.1.10) will not change.

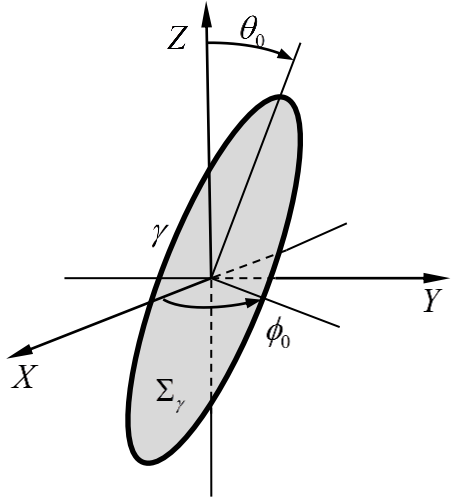

Fig.2 Integration path

One can see from (1.1.9) that integral over $d\phi$ is different from zero, and integral over $d\theta$ gives zero, therefore the calculation of $\operatorname{rot}\langle\vec{v}\rangle$ is analogue to the procedure described in our paper [32]. As a result, we obtain

$$\operatorname{rot}\langle\vec{v}\rangle = \gamma \operatorname{rot}\vec{A} = \gamma\vec{B} = -4\pi\alpha k \delta^2(\rho,\phi)\vec{e}_z,$$

$$\vec{B} = -\frac{2\alpha k}{\gamma}\frac{\delta(\rho)}{\rho}\vec{e}_z,$$

or

$$\vec{B} = -\frac{q_m^{(Wb)}\delta(\rho)}{2\pi\rho}\vec{e}_z, \qquad (1.1.11)$$

where

$$\delta^2(\rho,\phi) = \frac{\delta(\rho)}{2\pi\rho}, \quad \int_\Sigma \delta^2(\rho,\phi)d\sigma = \int_0^{2\pi} d\phi \int_0^R \frac{\delta(\rho)}{2\pi\rho}\rho d\rho = 1,$$

$$q_m^{(Wb)} = \frac{2\pi\hbar k}{q_e},$$

Value $q_m^{(Wb)}$ coincides with the known expression for the magnetic charge and satisfies the quantization condition [28, 32].

*Remark 1*

*The system obtained (1.1.11) is not the Dirac monopole, though it has «charge», which is numerically equal to the magnetic charge $q_m^{(Wb)}$.*

*First, unlike the Dirac monopole, the «magnetic» charge $q_m^{(Wb)}$ in (1.1.11) has only **one** line of force coinciding with the OZ axis and $\operatorname{div}\vec{B} = 0$.*

*Second, the magnetic field of the Dirac monopole $\sim \frac{1}{r^2}$, and the magnetic field from expression (1.1.11) is $\sim \frac{1}{r}$, which corresponds to the classical behavior of the magnetic field.*

If there is a set $N$ of such «magnetic» charges $q_m^{(Wb)}$, which centers are located, for example, in one plane with coordinate $\vec{\rho}_i$, $i=1,...,N$, then total magnetic field $\vec{B}$ produced by such a system may be formally represented, according to (1.1.11) in the following form:

$$\vec{B}(\vec{\rho}) = -\sum_{i=1}^{N}\frac{q_m^{(Wb)}\delta(|\vec{\rho}-\vec{\rho}_i|)}{2\pi|\vec{\rho}-\vec{\rho}_i|}\vec{e}_z. \qquad (1.1.12)$$

Thus, field $\langle\vec{v}\rangle$ has sources $Q$ and $\vec{B}$, which are different from zero in general case. From (1.1.7) it follows that the sources of irrotational field $Q$ exist, if $n \neq 0$, and from (1.1.11) it is followed that the sources of solenoidal field $\vec{B}$ exist, if $k \neq 0$. Consequently, the first term $n\theta$ in (1.1.5) is responsible for irrotational components and the second term $k\phi$ is responsible for the solenoidal components of the probability flow $\langle\vec{v}\rangle$.

Subject to the above mentioned, one can use the following expression for velocity $\langle \vec{v} \rangle$ according to the Helmholtz decomposition:

$$\langle \vec{v} \rangle = -\alpha \nabla \Phi + \gamma \vec{A} = -2\alpha \frac{n}{r} \vec{e}_\theta - \frac{2\alpha k}{r \sin \theta} \vec{e}_\phi,$$

where

$$\Phi = 2\varphi, \quad \varphi = n\theta + c_0, \quad \vec{A} = -\frac{2\alpha}{\gamma} \frac{k}{r \sin \theta} \vec{e}_\phi, \quad (1.1.13)$$

where $c_0$ is a certain coordinate-independent function. With that, the conditions $\operatorname{rot} \nabla \Phi = 0$, $\operatorname{div} \vec{A} = 0$ are met according to (1.1.7) and (1.1.11).

According to (1.1.11), the expression for the solenoidal field $\vec{A}$ may be rewritten in the following form:

$$\vec{A} = -\frac{q_m^{(Wb)}}{2\pi \rho} \vec{e}_\phi, \quad \oint_{\gamma_\rho} \vec{A} d\vec{l} = -q_m^{(Wb)}$$

*Remark 2*

*A certain function depending on time only may be chosen as function $c_0$, for example, $c_0(t) = -\omega t$, where $\omega = \mathrm{E}/\hbar$. As it is shown below, value $\mathrm{E}$ corresponds to the total energy. In case of non-conservative systems, the total energy may depend on time, therefore the linear dependence $c_0(t) = -\frac{\mathrm{E}}{\hbar} t$ is just a particular case of function $c_0(t)$.*

Let us find probability density function $f(\vec{r},t)$, corresponding to vector field (1.1.6). Using (1.1.6), (1.1.7) in (1.1.2), we obtain the equation for finding $f(\vec{r},t)$

$$\frac{\partial S}{\partial t} - 2\alpha \left( \frac{k}{r^2 \sin^2 \theta} \frac{\partial S}{\partial \phi} + \frac{n}{r^2} \frac{\partial S}{\partial \theta} \right) = -\frac{2\alpha n}{r^2} \operatorname{ctg} \theta,$$

$$-\frac{r^2}{2\alpha} \frac{\partial S}{\partial t} + \frac{k}{\sin^2 \theta} \frac{\partial S}{\partial \phi} + n \frac{\partial S}{\partial \theta} = n \operatorname{ctg} \theta, \quad (1.1.14)$$

Equation (1.1.14) is a linear one heterogeneous in partial differential relative to function $S(r,\phi,\theta,t)$, therefore we will seek complementary solution (1.1.14) in the form of the superposition of general homogeneous solution $S_{g.hom.}$ and particular heterogeneous solution $S_{p.het.}$:

$$S_{g.het.} = S_{g.hom.} + S_{p.het.}. \quad (1.1.15)$$

We will seek solution $S_{p.het.}$ in the form of $S_{p.het.} = S_{p.het.}(\theta)$, then equation (1.1.14) will have the following form:

$$\frac{\partial S_{p.het.}}{\partial \theta} = \text{ctg}\,\theta,$$

$$S_{p.het.} = \int \text{ctg}\,\theta\, d\theta = \ln|\sin\theta| + c,$$

$$S_{p.het.} = \ln(c_1 \sin\theta). \qquad (1.1.16)$$

Let us find solution $S_{g.hom.}$, which must satisfy the homogeneous equation

$$-\frac{r^2}{2\alpha}\frac{\partial S_{g.hom.}}{\partial t} + n\frac{\partial S_{g.hom.}}{\partial \theta} + \frac{k}{\sin^2\theta}\frac{\partial S_{g.hom.}}{\partial \phi} = 0. \qquad (1.1.17)$$

Equation (1.1.17) is a hyperbolic equation; consequently, it allows solution in the form of characteristics. As equation (1.1.17) contains three variables $t, \theta, \phi$, then there are two integrals of motion $\xi = c_1$ and $\eta = c_2$; the following condition must be met along these integrals:

$$S_{g.hom.}(\xi, \eta) = const. \qquad (1.1.18)$$

Introducing the parameterization $\theta = \theta(\tau)$, $\phi = \phi(\tau)$, $t = t(\tau)$ for total differential from $S_{g.hom.}$ along characteristics $(\xi, \eta)$ according to (1.1.18), we obtain

$$\frac{dS_{g.hom.}}{d\tau} = \frac{\partial S_{g.hom.}}{\partial t}\frac{dt}{d\tau} + \frac{\partial S_{g.hom.}}{\partial \theta}\frac{d\theta}{d\tau} + \frac{\partial S_{g.hom.}}{\partial \phi}\frac{d\phi}{d\tau} = 0. \qquad (1.1.19)$$

Comparing (1.1.19) and (1.1.17), we obtain

$$\frac{dt}{d\tau} = -\frac{r^2}{2\alpha},\ \frac{d\theta}{d\tau} = n,\ \frac{d\phi}{d\tau} = \frac{k}{\sin^2\theta}, \qquad (1.1.20)$$

which leads to characteristic equations

$$\frac{d\theta}{d\phi} = \frac{n}{k}\sin^2\theta, \qquad \frac{d\theta}{dt} = -\frac{2\alpha n}{r^2}. \qquad (1.1.21)$$

The solutions of characteristic equations (1.1.21) have the following form:

$$\int\frac{d\theta}{\sin^2\theta} = \frac{n}{k}\int d\phi, \qquad \int d\theta = -\frac{2\alpha n}{r^2}\int dt,$$

$$\phi = -\frac{k}{n}\text{ctg}\,\theta + c_1, \qquad \theta = -\frac{2\alpha n}{r^2}t + c_2, \qquad (1.1.22)$$

where $c_1, c_2$ are constant values. Expressions (1.1.22) determine the equations of motion integrals; value $S_{g.hom.}$ remains constant along these integrals (1.1.18)

$$\xi(\theta,\phi) = \phi + \frac{k}{n}\operatorname{ctg}\theta, \qquad (1.1.23)$$

$$\eta(r,\theta,t) = \theta + \frac{2\alpha n}{r^2}t.$$

Notice that there are no derivatives with respect to $r$ in equation (1.1.17), though solution $S_{g.hom.}$ itself may depend on $r$. If $S_{g.hom.} = S_{g.hom.}(r,\phi,\theta,t)$, then, in view of (1.1.23), the solution of equation (1.1.17) may be expressed in the following form

$$S_{g.hom.}(r,\theta,\phi,t) = G(r,\xi,\eta) = G\left(r, \phi + \frac{k}{n}\operatorname{ctg}\theta, \theta + \frac{2\alpha n}{r^2}t\right), \qquad (1.1.24)$$

where $G(r,\xi,\eta)$ is a certain differentiable function found from initial or boundary conditions. If the initial condition is established as follows

$$S_{g.hom.}(r,\phi,\theta,t)\big|_{t=0} = G_0(r,\phi,\theta), \qquad (1.1.25)$$

from (1.1.25), taking into account (1.1.24), it follows

$$G_0(r,\phi,\theta) = G\left(r, \phi + \frac{k}{n}\operatorname{ctg}\theta, \theta\right) = G(r,\tilde{\xi},\tilde{\eta}), \qquad (1.1.26)$$

where it is assigned

$$\tilde{\xi} \stackrel{\text{det}}{=} \phi + \frac{k}{n}\operatorname{ctg}\theta, \qquad \tilde{\eta} \stackrel{\text{det}}{=} \theta,$$

it follows from here that

$$\theta = \tilde{\eta}, \qquad \phi = \tilde{\xi} - \frac{k}{n}\operatorname{ctg}\tilde{\eta}. \qquad (1.1.27)$$

Inserting (1.1.27) into (1.1.26), we obtain

$$G(r,\tilde{\xi},\tilde{\eta}) = G_0\left(r, \tilde{\xi} - \frac{k}{n}\operatorname{ctg}\tilde{\eta}, \tilde{\eta}\right). \qquad (1.1.28)$$

Thus, on account of (1.1.28) and (1.1.23), solution (1.1.24) satisfying initial condition (1.1.25) has the following form:

$$S_{g.hom.}(r,\phi,\theta,t) = G_0\left(r, \xi - \frac{k}{n}\operatorname{ctg}\eta, \eta\right) = G_0\left(r, \phi + \frac{k}{n}\left(\operatorname{ctg}\theta - \operatorname{ctg}\left(\theta + \frac{2\alpha n}{r^2}t\right)\right), \theta + \frac{2\alpha n}{r^2}t\right).$$

(1.1.29)

In view of (1.1.15), (1.1.16) and (1.1.29) the complementary solution of equation (1.1.14) has the following form:

$$S_{g.het.}(r,\phi,\theta,t) = G_0\left(r, \phi + \frac{k}{n}\left(\operatorname{ctg}\theta - \operatorname{ctg}\left(\theta + \frac{2\alpha n}{r^2}t\right)\right), \theta + \frac{2\alpha n}{r^2}t\right) + \ln\sin\theta. \quad (1.1.30)$$

Unlike (1.1.29), solution (1.1.30) is not constant along characteristics (1.1.23), because (1.1.30) contains addend $S_{p.het.}$. Thus, the non-homogeneity of (1.1.14) results in the change of $S_{g.het.}$ along the characteristics. On the other hand, the heterogeneity of equation (1.1.14) is caused by the presence of sources $Q$ of the irrotational field (1.1.2), (1.1.7). If $Q=0$, for example, at $n=0$, then the original problem will be reduced to the problem considered in article [32].

As a result, for the probability density function $f(r,\phi,\theta,t)$ we obtain:

$$f(r,\phi,\theta,t) = e^{-S_{g.het.}(r,\phi,\theta,t)} = \frac{1}{\sin\theta}\exp\left[-G_0\left(r,\phi + \frac{k}{n}\left(\operatorname{ctg}\theta - \operatorname{ctg}\left(\theta + \frac{2\alpha n}{r^2}t\right)\right), \theta + \frac{2\alpha n}{r^2}t\right)\right],$$

or

$$f(r,\phi,\theta,t) = \frac{1}{\sin\theta} F_0\left(r, \phi + \frac{k}{n}\left(\operatorname{ctg}\theta - \operatorname{ctg}\left(\theta + \frac{2\alpha n}{r^2}t\right)\right), \theta + \frac{2\alpha n}{r^2}t\right), \quad (1.1.31)$$

where $F_0 \stackrel{det}{=} e^{-G_0}$. With that, function $F_0$ is determined from the initial conditions:

$$\left. f(r,\phi,\theta,t)\right|_{t=0} = \frac{F_0(r,\phi,\theta)}{\sin\theta} = f_0(r,\phi,\theta),$$
$$F_0(r,\phi,\theta) = \sin\theta f_0(r,\phi,\theta),$$

(1.1.32)

where $f_0(r,\phi,\theta)$ is known function of the initial distribution of the probability density.

### Remark 3
*Obtained solution (1.1.31) is periodic with respect to the second argument on the sphere. The second argument*

$$\phi + \frac{k}{n}\left[\operatorname{ctg}\theta - \operatorname{ctg}\left(\theta + \frac{2\alpha n}{r^2}t\right)\right] = \phi + \frac{k}{n}\left[\operatorname{ctg}\theta - \operatorname{ctg}(\theta + \omega_n t)\right],$$

*contains the addend $\omega_n t$, where*

$$\omega_n(r) \overset{\text{det}}{=} \frac{2\alpha n}{r^2} = -\frac{\hbar n}{mr^2}.$$

*Frequency $\omega_n$ is constant on the sphere, that is with a fixed radius $r$. As function ctg is periodic with the period $T_1 = \pi$, then solution (1.1.31) will be periodic on the sphere with respect to the second argument with the period $T_1$*

$$T_1 \overset{\text{det}}{=} \frac{m}{\hbar n} T_1 r^2 = \frac{m}{\hbar n} \pi r^2 \overset{\text{det}}{=} \frac{m}{\hbar n} \frac{\sigma}{4},$$

where $\sigma = 4\pi r^2$ is area of the sphere with a radius $r$.

*If the third argument of solution (1.1.31) is an argument of a certain periodic function of a period $T_2$, then solution (1.1.31) will have the second period $T_2$ on the sphere*

$$T_2 \overset{\text{det}}{=} \frac{m}{\hbar n} T_2 r^2.$$

*Having two periods $T_1$ and $T_2$, one can consider solution (1.1.31) formally as elliptic function on the sphere.*

*Period $T_1$ depends on the sphere radius $r$ and increases with the radius increasing. Considering concentric spheres, we obtain that in the origin of coordinates at $r \to 0$ period $T_1 \to 0$ and frequency $\omega_n \to \infty$ on ones with bigger radius $r \to \infty$ period $T_1 \to \infty$ and frequency $\omega_n \to 0$.*

*Notice that one can meet a similar analogue when comparing the micro and macro worlds. Periods of oscillators are small ($\sim 10^{-8}$ s) in the micro world (atom), and periods are big ($\sim 1$ year) in the macro world (solar system).*

Let us show that probability flow vector field $\langle \vec{v} \rangle$ (1.1.6) is tangential to characteristics (1.1.22). Suppose $\vec{u}$ is tangential vector along characteristic (1.1.22), then the components $\vec{u}$ having the following form according to (1.1.20):

$$x = \sin\theta \cos\phi, \ y = \sin\theta \sin\phi, \ z = \cos\theta,$$

$$\vec{u} = \begin{pmatrix} x'_\tau \\ y'_\tau \\ z'_\tau \end{pmatrix} = \frac{1}{\sin\theta} \begin{pmatrix} n\sin\theta\cos\theta\cos\phi - k\sin\phi \\ n\sin\theta\cos\theta\sin\phi + k\cos\phi \\ -n\sin^2\theta \end{pmatrix}. \tag{1.1.33}$$

The vector field $\langle \vec{v} \rangle$ (1.1.6) in the Cartesian coordinate system is of the form:

$$\vec{e}_r = \begin{pmatrix} \sin\theta\cos\phi \\ \sin\theta\sin\phi \\ \cos\theta \end{pmatrix}, \quad \vec{e}_\theta = \begin{pmatrix} \cos\theta\cos\phi \\ \cos\theta\sin\phi \\ -\sin\theta \end{pmatrix}, \quad \vec{e}_\phi = \begin{pmatrix} -\sin\phi \\ \cos\phi \\ 0 \end{pmatrix}, \quad (1.1.34)$$

$$\langle\vec{v}\rangle = -\frac{2\alpha}{r\sin\theta}\left(k\vec{e}_\phi + n\sin\theta\vec{e}_\theta\right) = -\frac{2\alpha}{r\sin\theta}\begin{pmatrix} n\sin\theta\cos\theta\cos\phi - k\sin\phi \\ n\sin\theta\cos\theta\sin\phi + k\cos\phi \\ -n\sin^2\theta \end{pmatrix}. \quad (1.1.35)$$

Comparing (1.1.33) and (1.1.35), we obtain

$$\langle\vec{v}\rangle = -\frac{2\alpha}{r}\vec{u}, \quad (1.1.36)$$

which was to be proved.

Figs. 3-5 show diagrams of characteristics (1.1.22) for various values of $n$ and $k$. As the azimuth angle $\phi$ is defined with a precision up to $2\pi$, constant value $c_1$ in expression (1.1.22) should be taken in the region $0 \leq c_1 < 2\pi$. Six characteristics are shown in each figure corresponding to six various values of constant $c_1^i, i = 0,...,5$ (1.1.22)

$$c_1^i = \frac{2\pi}{6}i \quad (1.1.37)$$

Figs. 3-5 show that the characteristics «connect» two opposite poles of the sphere ($\theta = 0$ and $\theta = \pi$).

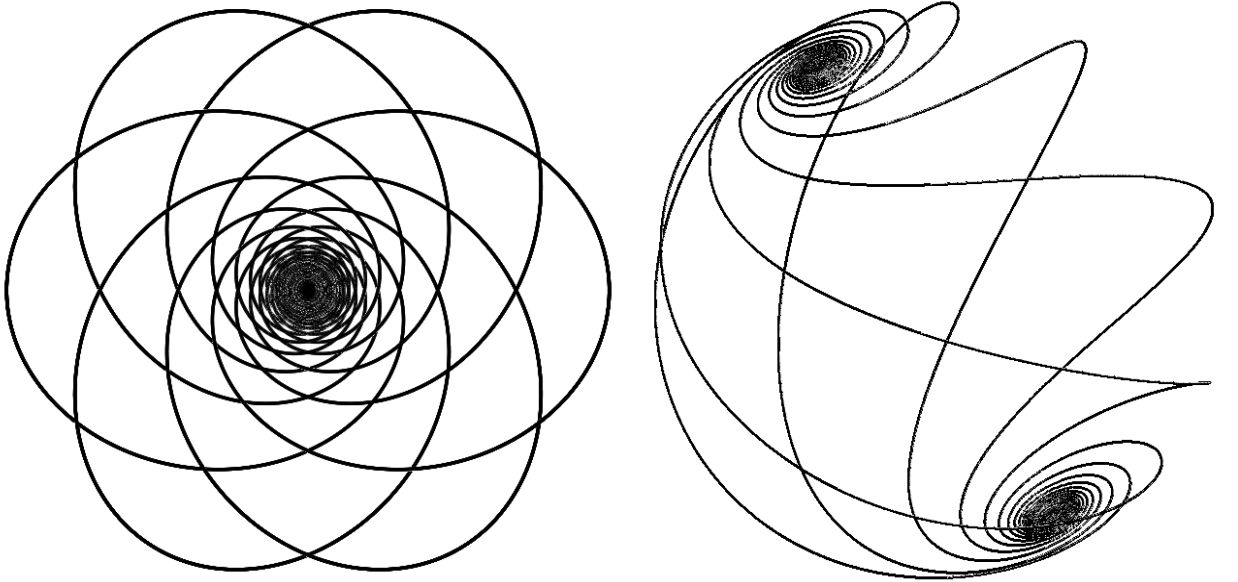

Fig.3 Diagram of the characteristics at n = 1 and k = 1

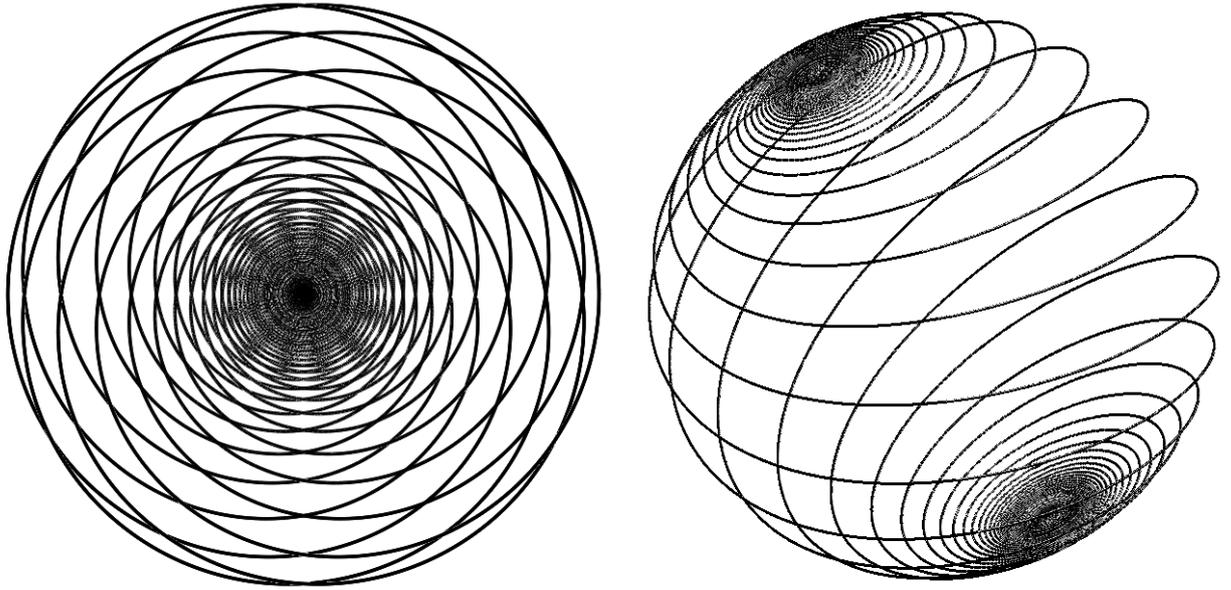

Fig.4 Diagrams of the characteristics at n = 1 and k = 3

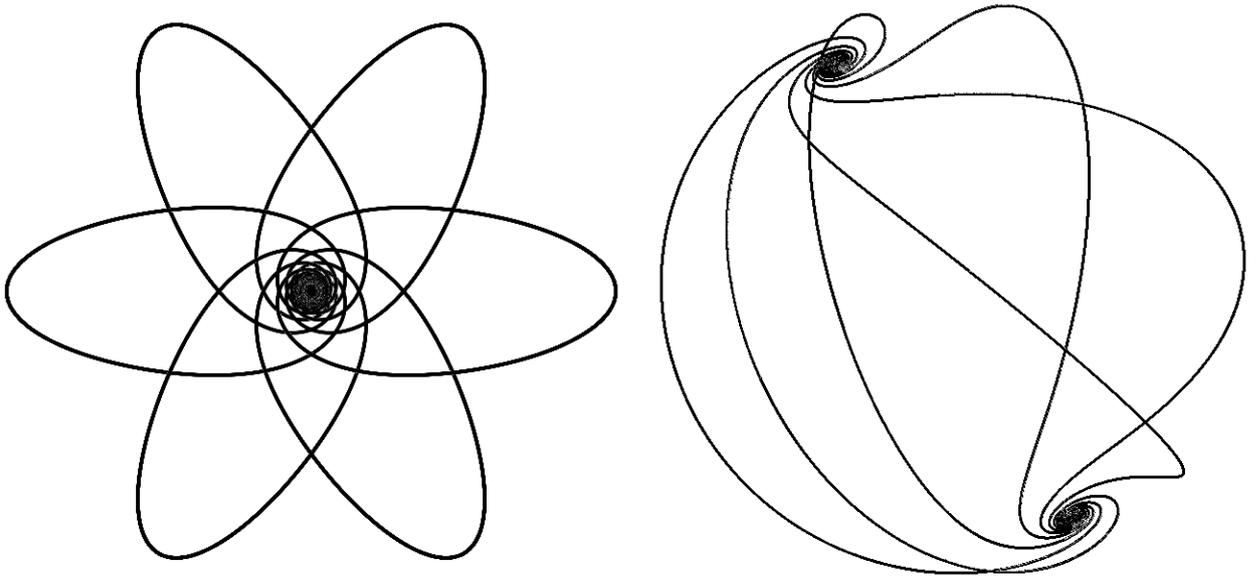

Fig.5 Diagrams of the characteristics at n = 3 and k = 1

***Remark 4***

*Notice that the characteristics behavior (1.1.22) in the pole area it seems leads to the existence shock wave compression and expansion one. That is by $z>0$ shock wave expansion ($Q>0$) takes place, and by $z<0$ shock wave compression ($Q<0$) takes place. However from (1.1.22) it follows that the characteristics do not reach the sphere poles (see Figs. 3-5), that is there is no end value of the azimuth angle $\phi_{end}$ for which $\theta=0$ or $\theta=\pi$.*

Probability density function (1.1.31) has a singularity in the sphere poles ($\theta=0$ and $\theta=\pi$) and may tend to infinity, therefore let us find its norm (total probability)

$$\int\limits_{(\infty)} f d\omega = \int\limits_0^{2\pi} d\phi \int\limits_0^{\pi} \sin\theta d\theta \int\limits_0^{+\infty} r^2 f dr =$$

$$= \int\limits_0^{2\pi} d\phi \int\limits_0^{\pi} d\theta \int\limits_0^{+\infty} r^2 F_0\left(r, \phi + \frac{k}{n}\left(\operatorname{ctg}\theta - \operatorname{ctg}\left(\theta + \frac{2\alpha n}{r^2}t\right)\right), \theta + \frac{2\alpha n}{r^2}t\right) dr. \qquad (1.1.38)$$

Thus, the norm $f$ function existence links with the $F_0$ function integral (1.1.38) existence. The $F_0$ function initial condition (1.1.32) is satisfied.

The expression for the wave function $\Psi$, according [27], (1.1.13), (1.1.31) and Remark 3, is as follows

$$\Psi(\vec{r},t) = \frac{1}{\sqrt{\sin\theta}} \sqrt{F_0\left(r, \phi + \frac{k}{n}\left[\operatorname{ctg}\theta - \operatorname{ctg}(\theta + \omega_n(r)t)\right], \theta + \omega_n(r)t\right)} e^{i\left(n\theta - \frac{E}{\hbar}t\right)}, \qquad (1.1.39)$$

Let us find corresponding potential $U$ for wave function (1.1.39). In view of [27] and (1.1.13), we obtain

$$U(\vec{r},t) = -\frac{1}{\beta}\left\{\frac{\partial \varphi(\vec{r},t)}{\partial t} + \alpha\left[\frac{\Delta\sqrt{f(\vec{r},t)}}{\sqrt{f(\vec{r},t)}} - |\nabla\varphi(\vec{r},t)|^2\right] + \gamma(\vec{A}, \nabla\varphi)\right\},$$

$$U(\vec{r},t) = -\frac{1}{\beta}\frac{\partial\varphi}{\partial t} + \frac{\alpha}{\beta}\left[|\nabla\varphi|^2 - \frac{\Delta|\Psi|}{|\Psi|}\right],$$

or

$$U = -\frac{1}{\beta}\frac{\partial\varphi}{\partial t} + \frac{\alpha}{\beta}|\nabla\varphi|^2 - Q, \qquad (1.1.40)$$

where the expression

$$Q \stackrel{\text{det}}{=} \frac{\alpha}{\beta}\frac{\Delta|\Psi|}{|\Psi|} = -\frac{\hbar^2}{2m}\frac{\Delta|\Psi|}{|\Psi|} \qquad (1.1.41)$$

is known as quantum potential in the pilot-wave theory [29-31]. The quantum potential is associated with classical potential $e\chi$ by the formula [27]

$$\chi = \frac{2\alpha\beta}{\gamma}\left(\frac{1}{2\alpha\beta}\frac{|\gamma\vec{A}|^2}{2} + U + Q\right). \qquad (1.1.42)$$

Inserting (1.1.13) and (1.1.40) into (1.1.42), we obtain the expression for the potential $e\chi$

$$e\chi = -\frac{m}{2}|\gamma\vec{A}|^2 + U + Q = -\frac{m}{2}|\gamma\vec{A}|^2 - \frac{\hbar^2}{2m}|\nabla\varphi|^2 + E = -\frac{(\hbar k)^2}{2m\rho^2} - \frac{(\hbar n)^2}{2mr^2} + E,$$

$$-e\chi = \frac{M_\phi^2}{2m\rho^2} + \frac{M_\theta^2}{2mr^2} - E = \frac{P_\phi^2}{2m} + \frac{P_\theta^2}{2m} - E = \frac{P^2}{2m} - E, \qquad (1.1.43)$$

$$E = \frac{P^2}{2m} + e\chi = const,$$

where

$$M_\theta \stackrel{det}{=} \hbar n, \quad M_\phi \stackrel{det}{=} \hbar k. \qquad (1.1.44)$$

Value $M_\theta$ may be interpreted as an angular moment of momentum associated with polar angle $\theta$ referred to the origin of coordinates, and value $M_\phi$ may be interpreted as an angular moment of momentum associated with azimuth angle $\phi$ referred to the OZ axis.

From expression (1.1.44) it follows that moments $M_\theta$ and $M_\phi$ are quantized.

The wave function must satisfy, according to [27], the Schrödinger equation of the form:

$$\frac{i}{\beta}\frac{\partial\Psi}{\partial t} = -\alpha\beta\left(\hat{p} - \frac{\gamma}{2\alpha\beta}\vec{A}\right)^2\Psi + \frac{1}{2\alpha\beta}\frac{|\gamma\vec{A}|^2}{2}\Psi + U\Psi.$$

Taking into account (1.1.43), we obtain

$$i\hbar\frac{\partial\Psi}{\partial t} = \frac{1}{2m}\left(\hat{p} - e\vec{A}\right)^2\Psi + (e\chi - Q)\Psi. \qquad (1.1.45)$$

At $k = 0$ according to (1.1.13), (1.1.40), equation (1.1.45) transits in the classical Schrödinger equation for a scalar particle

$$i\hbar\frac{\partial\Psi}{\partial t} = \frac{\hat{p}^2}{2m}\Psi + U\Psi. \qquad (1.1.46)$$

Classical motion equations, according to [27], are as follows

$$\frac{d\langle\vec{v}\rangle}{dt} = -\gamma\left(\vec{E} + [\langle\vec{v}\rangle, \vec{B}]\right), \qquad (1.1.47)$$

where

$$\vec{E} \stackrel{det}{=} -\nabla\chi - \frac{\partial\vec{A}}{\partial t}, \quad \vec{B} = \operatorname{rot}\vec{A}.$$

Let us show that (1.1.47) is exercised for the velocity field $\langle\vec{v}\rangle$ (1.1.13) and potential $\chi$ (1.1.43). Using (1.1.31), let us calculate $\frac{d\langle\vec{v}\rangle}{d\tau}$. Value $t$ is physical time and value $\tau$ in (1.1.20) is used for parameterization of characteristic (1.1.22), therefore

$$\frac{d\langle \vec{v}\rangle}{dt} = \frac{d\langle \vec{v}\rangle}{d\tau}\frac{d\tau}{dt}, \qquad (1.1.48)$$

Having found $\dfrac{d\langle \vec{v}\rangle}{d\tau}$, we obtain

$$\frac{d\langle \vec{v}\rangle}{d\tau} = -\frac{2\alpha}{r}\frac{d}{d\tau}\begin{pmatrix} n\cos\theta\cos\phi - k\dfrac{\sin\phi}{\sin\theta} \\ n\cos\theta\sin\phi + k\dfrac{\cos\phi}{\sin\theta} \\ -n\sin\theta \end{pmatrix} =$$

$$= -\frac{2\alpha}{r}\begin{pmatrix} -n^2\sin\theta\cos\phi - nk\dfrac{\cos\theta\sin\phi}{\sin^2\theta} - k\dfrac{k\cos\phi - n\sin\phi\cos\theta\sin\theta}{\sin^3\theta} \\ -n^2\sin\theta\sin\phi + nk\dfrac{\cos\theta\cos\phi}{\sin^2\theta} + k\dfrac{-k\sin\phi - n\cos\phi\cos\theta\sin\theta}{\sin^3\theta} \\ -n^2\cos\theta \end{pmatrix}.$$

As a result,

$$\frac{d\langle \vec{v}\rangle}{d\tau} = \frac{2\alpha}{r\sin^3\theta}\begin{pmatrix} n^2\sin^4\theta\cos\phi + k^2\cos\phi \\ n^2\sin^4\theta\sin\phi + k^2\sin\phi \\ n^2\cos\phi\sin^3\theta \end{pmatrix}. \qquad (1.1.49)$$

From expression (1.1.20) it follows that

$$\frac{d\tau}{dt} = -\frac{2\alpha}{r^2} = \frac{\hbar}{mr^2}. \qquad (1.1.50)$$

Inserting (1.1.49), (1.1.50) into (1.1.48), we obtain the expression for the acceleration $\dfrac{d\langle \vec{v}\rangle}{dt}$

$$\frac{d\langle \vec{v}\rangle}{dt} = -\frac{\hbar^2}{m^2 r^3 \sin^3\theta}\begin{pmatrix} n^2\sin^4\theta\cos\phi + k^2\cos\phi \\ n^2\sin^4\theta\sin\phi + k^2\sin\phi \\ n^2\cos\phi\sin^3\theta \end{pmatrix}. \qquad (1.1.51)$$

Let us calculate the right side of equation (1.1.47). According to (1.1.43), we obtain

$$-e\nabla\chi = -\frac{\hbar^2}{mr^3}\left(n^2 + \frac{k^2}{\sin^2\theta}\right)\vec{e}_r - \frac{\hbar^2 k^2}{mr^3}\frac{\cos\theta}{\sin^3\theta}\vec{e}_\theta.$$

Converting into the Cartesian coordinate system, we obtain, according to (1.1.34),

$$-e\nabla\chi = -\frac{\hbar^2}{mr^3}\begin{pmatrix}\left(n^2+\dfrac{k^2}{\sin^2\theta}\right)\sin\theta\cos\phi+\dfrac{k^2\cos^2\theta\cos\phi}{\sin^3\theta}\\ \left(n^2+\dfrac{k^2}{\sin^2\theta}\right)\sin\theta\sin\phi+\dfrac{k^2\cos^2\theta\sin\phi}{\sin^3\theta}\\ \left(n^2+\dfrac{k^2}{\sin^2\theta}\right)\cos\theta-\dfrac{k^2\cos\theta\sin\phi}{\sin^3\theta}\end{pmatrix},$$

$$-e\nabla\chi = -\frac{\hbar^2}{mr^3\sin^3\theta}\begin{pmatrix}n^2\sin^4\theta\cos\phi+k^2\cos\phi\\ n^2\sin^4\theta\sin\phi+k^2\sin\phi\\ n^2\cos\phi\sin^3\theta\end{pmatrix}. \qquad (1.1.52)$$

In view of (1.1.11), field $\vec{B}$ is different from zero only on the OZ axis, that is on the sphere poles. From Remark 4 it follows that the trajectory (characteristics) does not reach the sphere poles, consequently

$$\left[\langle\vec{v}\rangle,\vec{B}\right]=\vec{0} \qquad (1.1.53)$$

Inserting (1.1.51), (1.1.52) and (1.1.53) into equation of motion (1.1.47), we obtain a valid identical equation, which was to be proved.

### Remark 5

*One can find classical potential $\chi$ (1.1.42), (1.1.43) assigning the velocities vector field $\langle\vec{v}\rangle$ in continuum classical mechanics (probability flow in case of quantum mechanics) according to motion equation (1.1.47). Thus, the distribution of potential $\chi$ is rigidly associated with the distribution of velocity field $\langle\vec{v}\rangle$. At that, there is an infinite aggregate of trajectories (characteristics) in the continuum satisfying motion equation (1.1.47) with various initial data (Cauchy problem). Calculating the total energy for a certain trajectory according to (1.1.43) and equation of Hamilton-Jacobi [27], we obtain*

$$-\frac{\partial\Phi}{\partial t}=\frac{2}{\hbar}W=\frac{2}{\hbar}\left(\frac{m}{2}|\langle\vec{v}\rangle|^2+e\chi\right)=\mathrm{E}. \qquad (1.1.54)$$

*The functions of kinetic and potential energy distribution in (1.1.54) have the same distribution law according to (1.1.35) and (1.1.43)*

$$\frac{m}{2}|\langle\vec{v}\rangle|^2=\frac{\hbar^2}{2mr^2}\left(n^2+\frac{k^2}{\sin^2\theta}\right), \qquad (1.1.55)$$

$$e\chi=-\frac{\hbar^2}{2mr^2}\left(n^2+\frac{k^2}{\sin^2\theta}\right)+\mathrm{E}.$$

*Compatibility (1.1.55) of velocity field $\langle\vec{v}\rangle$ and potential $\chi$ is typical of classic continuum mechanics. Compatibility (1.1.55) yields the presence of the continuum of trajectories and energies* E. *Such distributions we will call self-consistent ones [27].*

*Self-consistency (1.1.55) is violated at transition to quantum mechanics.*

*In § 2.1 the motion in a modified Coulomb potential field as an example of the violation of the self-consistency (1.1.55) was considered*

## 1.2 Intrinsic magnetic moment

Let us find the intrinsic magnetic moment of the considered full-sphere. The intrinsic magnetic moment of the full-sphere is produced by the density of electric current

$$\vec{J} \stackrel{det}{=} ef \langle \vec{v} \rangle, \qquad (1.2.1)$$

flowing along the characteristics with flow rates $\langle \vec{v} \rangle$ (1.1.35), (1.1.36). The magnetic moment of the region $\Omega$ can be calculated with the following formula

$$\vec{\mu}_s = \frac{1}{2} \int_\Omega [\vec{r}, \vec{J}] d\omega. \qquad (1.2.2)$$

Inserting the expression for the velocity $\langle \vec{v} \rangle$ (1.1.35) into (1.2.1), we obtain

$$\vec{J} = ef \langle \vec{v} \rangle = f \frac{e\hbar}{m}\left(\frac{n}{r}\vec{e}_\theta + \frac{k}{r\sin\theta}\vec{e}_\phi\right) = 2\mu_B f\left(\frac{n}{r}\vec{e}_\theta + \frac{k}{r\sin\theta}\vec{e}_\phi\right), \qquad (1.2.3)$$

where $\mu_B \stackrel{det}{=} \frac{e\hbar}{2m}$ is Bohr magneton. Let us consider the simplest model of the density flow $f = f(r)$. Notice that $S(r) = -\ln f(r)$ is the solution of the equation (1.1.14) at $n=0$.

Converting into the spherical coordinate system and taking into account (1.2.3), we obtain the expression for the intrinsic magnetic moment of the full-sphere (1.2.2) of the following form

$$\begin{aligned}\vec{\mu}_s &= \frac{1}{2}\int_0^{2\pi} d\phi \int_0^\pi r^2\sin\theta d\theta \int_0^{+\infty} 2\mu_B rf(r)\left(\frac{k}{r\sin\theta}[\vec{e}_r,\vec{e}_\phi] + \frac{n}{r}[\vec{e}_r,\vec{e}_\theta]\right)dr = \\ &= \mu_B \int_0^{2\pi} d\phi \int_0^\pi \sin\theta d\theta \int_0^{+\infty} r^2 f(r)\left(\frac{k}{\sin\theta}\vec{e}_\theta + n\vec{e}_\phi\right)dr = \\ &= \mu_B k \int_0^{2\pi} d\phi \int_0^\pi \vec{e}_\theta d\theta \int_0^{+\infty} r^2 f(r)dr + \mu_B n \int_0^{2\pi} d\phi \int_0^\pi \vec{e}_\phi \sin\theta d\theta \int_0^{+\infty} r^2 f(r)dr = \vec{I}_1 + \vec{I}_2. \end{aligned} \qquad (1.2.4)$$

Let us find the integrals $\vec{I}_1$ and $\vec{I}_2$ separately, having used the Cartesian expressions (1.1.34) for the spherical reference $\vec{e}_\theta$ and $\vec{e}_\phi$.

$$\vec{I}_1 \overset{det}{=} \mu_B k \int_0^{2\pi} d\phi \int_0^{\pi} \vec{e}_\theta d\theta \int_0^{+\infty} r^2 f(r) dr = \mu_B k \int_0^{2\pi} d\phi \int_0^{\pi} \begin{pmatrix} \cos\theta\cos\phi \\ \cos\theta\sin\phi \\ -\sin\theta \end{pmatrix} d\theta \int_0^{+\infty} r^2 f(r) dr =$$

$$= \mu_B k \begin{pmatrix} \int_0^{2\pi}\cos\phi d\phi \int_0^{\pi}\cos\theta d\theta \\ \int_0^{2\pi}\sin\phi d\phi \int_0^{\pi}\cos\theta d\theta \\ -\int_0^{2\pi} d\phi \int_0^{\pi}\sin\theta d\theta \end{pmatrix} \int_0^{+\infty} r^2 f(r) dr = \mu_B k \begin{pmatrix} 0 \\ 0 \\ 4\pi \end{pmatrix} \int_0^{+\infty} r^2 f(r) dr = \mu_B k 4\pi \int_0^{+\infty} r^2 f(r) dr \vec{e}_z,$$

as the function $f(r)$ satisfies the condition $4\pi \int_0^{+\infty} r^2 f(r) dr = 1$ due to normalizing, we obtain

$$\vec{I}_1 = \mu_B k \vec{e}_z. \tag{1.2.5}$$

$$\vec{I}_2 \overset{det}{=} \frac{\mu_B n}{V} \int_0^{2\pi} d\phi \int_0^{\pi} \vec{e}_\phi \sin\theta d\theta \int_0^{+\infty} r^2 f(r) dr = \mu_B n \int_0^{2\pi} \begin{pmatrix} -\sin\phi \\ \cos\phi \\ 0 \end{pmatrix} d\phi \int_0^{\pi} \sin\theta d\theta \int_0^{+\infty} r^2 f(r) dr = \mu_B n \begin{pmatrix} 0 \\ 0 \\ 0 \end{pmatrix},$$

$$\vec{I}_2 = \vec{0}. \tag{1.2.6}$$

Inserting (1.2.5) and (1.2.6) into (1.2.4), we obtain the expression for the intrinsic magnetic moment of the full-sphere

$$\vec{\mu}_s = \mu_B k \vec{e}_z. \tag{1.2.7}$$

At $k = \pm 1$ (see Figs. 3-5) expression (1.2.7) coincides with the intrinsic magnetic moment of an electron with the spin $s = \pm \frac{1}{2}$ and $g$-factor, $g = 2$

$$\mu_s = s\,\mu_B\,g = \pm\mu_B. \tag{1.2.7*}$$

The intrinsic magnetic moment $\vec{\mu}_s$ value is found from classical continuum mechanics concepts using the solenoidal $\langle \vec{v} \rangle$ (1.1.6) and is interpreted as rotating motion, i.e. spin.

As it is known, there was a hypothesis in the early stages of quantum mechanics development that spin is associated with the electron axial rotation. However, at evaluating the tangential velocity, which «the electron surface» should rotate, a superlight value was obtained [40-42].

In the considered case, the velocity $\langle \vec{v} \rangle$ has a pole on the OZ axis, but this pole is not reached by the characteristics due to Remark 3 and $n = 0$ (for $f = f(r)$).

***Let us show that the problem of the rotation velocity $\langle v \rangle$ surpassing the speed of light $c$ can be solved for the velocity (1.1.6) by choosing the density function $f$.***

As the rotation about the OZ axis determines the preferred direction in space, it seems natural to consider not the spherically symmetric density distribution $f = f(r)$, but the cylindrical symmetric distribution $f = f(\rho, z)$ [43]. The function $S(\rho, z) = -\ln f(\rho, z)$ satisfies equation (1.1.2) due to (1.1.7) at $n = 0$.

Doing computations analogue to (1.2.4)-(1.2.7), we obtain

$$\vec{\mu}_s = \frac{1}{2} \int_{-\infty}^{+\infty} dz \int_0^{2\pi} \rho d\phi \int_0^{+\infty} 2\mu_B r f(\rho, z) \frac{k}{r \sin \theta} \left[\vec{e}_r, \vec{e}_\phi\right] d\rho =$$

$$= \mu_B k \int_{-\infty}^{+\infty} dz \int_0^{2\pi} \rho d\phi \int_0^{+\infty} \frac{f(\rho, z)}{\sin \theta} \begin{pmatrix} \cos \theta \cos \phi \\ \cos \theta \sin \phi \\ -\sin \theta \end{pmatrix} d\rho =$$

$$= -2\pi \mu_B k \vec{e}_z \int_{-\infty}^{+\infty} dz \int_0^{+\infty} \rho \frac{f(\rho, z)}{\sin \theta} \sin \theta d\rho = -2\pi \mu_B k \vec{e}_z \int_{-\infty}^{+\infty} dz \int_0^{+\infty} \rho f(\rho, z) d\rho,$$

$$\vec{\mu}_s = \mu_B k \vec{e}_z, \tag{1.2.7*}$$

where, due to normalizing $f = f(\rho, z)$, the following is taken into account

$$2\pi \int_{-\infty}^{+\infty} dz \int_0^{+\infty} \rho f(\rho, z) d\rho = 1. \tag{1.2.8}$$

As a result, an expression is obtained for the cylindrically symmetric distribution $f = f(\rho, z)$ analogue to (1.2.7). In both cases — (1.2.7) and (1.2.7*) — the intrinsic magnetic moment $\vec{\mu}_s$ *does not depend* explicitly on the kind of the density function $f$. Notice that expression (1.2.7*) for the intrinsic magnetic moment $\vec{\mu}_s$ is obtained analytically rigorously on the basis of the expression of the velocity $\langle \vec{v} \rangle$ (1.1.6), which leads also to the principle of Bohr–Sommerfeld quantization at considering multivalent Riemannian surface [32]. At that there has been no need to modify $g$-factor, as it is done in the Dirac and Pauli equations.

In order for the flow velocity $\langle v \rangle$ not to surpass the speed of light $c$, it is required that the density function $f = f(\rho, z)$ compensates the pole $\langle v \rangle \sim \dfrac{1}{\rho}$. The following functions can be chosen for $f = f(\rho, z)$, which has such a property:

$$f_1(\rho, z) = C_1 \frac{1}{\rho^2} e^{-\frac{(\ln(\kappa \rho) - \mu)^2}{2\sigma^2}} e^{-\frac{z^2}{2\gamma^2}}, \tag{1.2.9}$$

$$f_2(\rho, z) = C_2 \rho^\alpha e^{-\beta \rho} e^{-\frac{z^2}{2\gamma^2}},$$

where $\alpha, \beta, \gamma, \kappa, \mu, \sigma$ are certain constant values, and $C_1, C_2$ are chosen from normalization condition (1.2.8)

$$1 = 2\pi C_1 \int_{-\infty}^{+\infty} e^{-\frac{z^2}{2\gamma^2}} dz \int_0^{+\infty} e^{-\frac{(\ln(\kappa\rho)-\mu)^2}{2\sigma^2}} \frac{d\rho}{\rho} = 2\pi\sqrt{2\pi}\gamma\sqrt{2\pi}\sigma C_1 = 4\pi^2 \gamma\sigma C_1,$$

$$1 = 2\pi C_2 \int_{-\infty}^{+\infty} e^{-\frac{z^2}{2\gamma^2}} dz \int_0^{+\infty} \rho^{\alpha+1} e^{-\beta\rho} d\rho = 2\pi C_2 \frac{\Gamma(\alpha+2)}{\beta^{\alpha+2}} \sqrt{2\pi}\gamma,$$

$$C_1 = \frac{1}{4\pi^2 \gamma\sigma}, \quad (1.2.10)$$

$$C_2 = \frac{\beta^{\alpha+2}}{(2\pi)^{3/2} \gamma \Gamma(\alpha+2)},$$

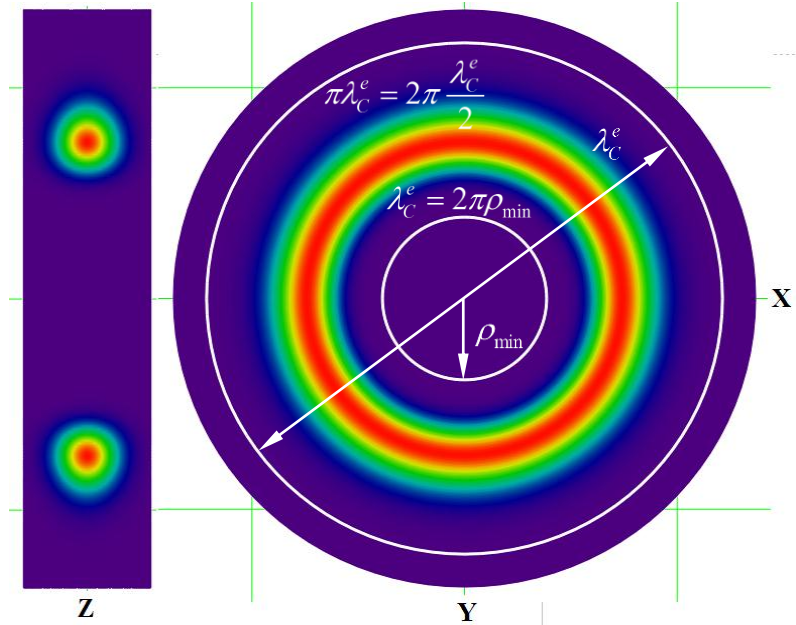

Fig. 6 Example of the density distribution $f_1 = f_1(\rho, z)$

Fig. 6 shows the density distribution $f_1(\rho, z)$ (1.2.9) at two sections – XOZ and XOY. The distribution $f_1(\rho, z)$ (1.2.9) is azimuthally symmetric and has a form of a «toroidal ring» (see Fig.6). As it is seen in Fig.6, the density distribution $f_1$ is focused inside the toroidal ring, and on the OZ axis, the density equals to zero. Consequently, choosing the values $\kappa, \mu, \sigma$, one may achieve the freedom from velocities surpassing the speed of light, let us demonstrate it.

The minimal radius $\rho_{min}$, starting with which $\rho > \rho_{min}$ the velocity $\langle v \rangle(\rho, \phi) < c$, according to (1.1.13), is of the following form

$$\langle v \rangle = \gamma A = \frac{\hbar}{m\rho_{min}} = c,$$

$$\rho_{min} = \frac{\hbar}{mc}. \quad (1.2.11)$$

Numerical value is $\rho_{min} \approx 3.862 \cdot 10^{-13} m$. Notice that Compton length of the electron $\lambda_C^e$ has the following value

$$\lambda_C^e = \frac{2\pi\hbar}{mc} = 2\pi\rho_{min},$$

that is $\lambda_C^e$ equals to the length of the circumference of a circle with the minimal radius $\rho_{min}$. Fig.7 shows the density distribution (1.2.9) in the median plane ($z = 0$) along the radius $\rho$.

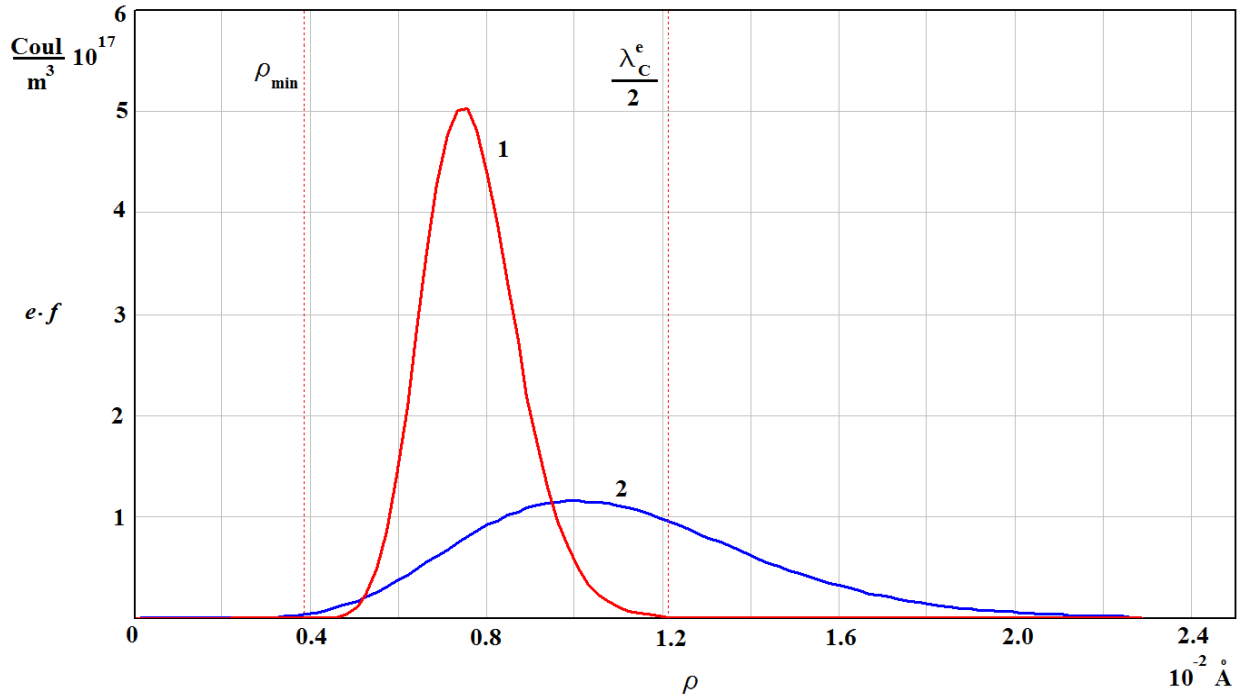

Fig. 7 Density distribution $f_1 = f_1(\rho, 0)$ и $f_2 = f_2(\rho, 0)$ along the radius $\rho$

The constant values $\alpha, \beta, \gamma, \kappa, \mu, \sigma$ in Fig.7 have values: $\alpha = 10$, $\frac{1}{\beta} = 100\,\text{fm}$, $\gamma = 100\,\text{fm}$, $\kappa = \pi/\lambda_C^e$, $\mu = 0$, $\sigma = 0.14$ The values of the magnitudes $\alpha, \beta, \gamma, \kappa, \mu, \sigma$ may vary from the chosen. The variation of the values $\alpha, \beta$ results in change in the maximal charge $e \cdot f$ density and its shift along the radius (for $f_2$), and the variation $\gamma^2$ determines the covariance «width» of the density distribution along the OZ axis OZ (for $f_1$) (see Fig. 6,7). The variation $\kappa$ shifts the peak of distribution $f_1$. As can be seen in Fig. 7, all velocities are less than the speed of light $\langle v \rangle < c$, as the density different from zero is located farther than the minimal radius $\rho_{min}$ (1.2.11).

*Thus, the considered «toroidal ring model» has the electrical charge, mass and intrinsic magnetic moment of the electron, that is its quantum numbers coincide with the ones of the electron, it is consistent with the relativity theory, therefore it can be considered as the **electron model**.*

The developed mathematical model of an «elementary» particle does not fit for the description of compound particles (consisting, for example, of quarks – the proton, neutron), but fits for the description of non-compound particles, such as the electron ($e$), muon ($\mu$) and tau ($\tau$).

### 1.3 Entropy of the model

The entropy $H$ of the random continued variable is of the following form by the convention

$$H(t) \stackrel{det}{=} -\int_{(\infty)} f(\vec{r},t)\ln f(\vec{r},t)d^3r, \qquad (1.3.1)$$

or taking into account designation $S = -\ln f$ (1.1.2)

$$H(t) = \int_{(\infty)} f(\vec{r},t)S(\vec{r},t)d^3r = N(t)\langle S\rangle(t), \qquad (1.3.2)$$

where

$$N(t) = \int_{(\infty)} f(\vec{r},t)d^3r.$$

If normalizing is performed for the total probability, then $N(t)=1$ and $H(t)=\langle S\rangle(t)$. From (1.3.1) and (1.3.2) it follows that the value $S(\vec{r},t)$ may be interpreted as the entropy at the point of space, and $H(t)$ determines an average entropy $\langle S\rangle$ through the entire space.

The entropy $S(\vec{r},t)$ satisfies equation (1.1.2), therefore we obtain the equation for $\langle S\rangle$ of the form (1.1.2). Let us multiply equation (1.1.1) by $(1+\ln f)$ and integrate over $d^3r$.

$$(1+\ln f)\frac{\partial f}{\partial t} + (1+\ln f)\operatorname{div}_r\left[\langle \vec{v}\rangle f\right] = 0,$$

$$\int_{(\infty)} \frac{\partial}{\partial t}(f \ln f)d^3r + \int_{(\infty)} (1+\ln f)\operatorname{div}_r\left[\langle \vec{v}\rangle f\right]d^3r = 0. \qquad (1.3.3)$$

The first integral in (1.3.3) is of the following form:

$$\int_{(\infty)} (1+\ln f)\frac{\partial f}{\partial t}d^3r = \int_{(\infty)} \frac{\partial}{\partial t}(f \ln f)d^3r = -\frac{\partial}{\partial t}\int_{(\infty)} fSd^3r = -\frac{d}{dt}\bigl[N(t)\langle S\rangle(t)\bigr]. \qquad (1.3.4)$$

The second integral in (1.3.1) is

$$\int_{(\infty)} (1+\ln f)\operatorname{div}_r\left[\langle \vec{v}\rangle f\right]d^3r = \int_{(\infty)} \bigl(\langle \vec{v}\rangle, (1+\ln f)\nabla_r f\bigr)d^3r + \int_{(\infty)} (1+\ln f)f\operatorname{div}_r\langle \vec{v}\rangle d^3r =$$

$$= \int_{(\infty)} \bigl(\langle \vec{v}\rangle, \nabla_r(f \ln f)\bigr)d^3r + \int_{(\infty)} f\operatorname{div}_r\langle \vec{v}\rangle d^3r + \int_{(\infty)} f\ln f\operatorname{div}_r\langle \vec{v}\rangle d^3r =$$

$$= -\int_{(\infty)} \bigl(\langle \vec{v}\rangle, \nabla_r(fS)\bigr)d^3r - \int_{(\infty)} fS\operatorname{div}_r\langle \vec{v}\rangle d^3r + \int_{(\infty)} f\operatorname{div}_r\langle \vec{v}\rangle d^3r =$$

$$= -\int_{(\infty)} \operatorname{div}_r\bigl[fS\langle \vec{v}\rangle\bigr]d^3r + \int_{(\infty)} f\operatorname{div}_r\langle \vec{v}\rangle d^3r = \int_{(\infty)} fQd^3r = N(t)\langle Q\rangle(t), \qquad (1.3.5)$$

where the condition $\int\limits_{(\infty)} \text{div}_r \left[ f S \langle \vec{v} \rangle \right] d^3 r = \int\limits_{\Sigma_\infty^r} f S \langle \vec{v} \rangle d\vec{\sigma}_r = 0$ is taken to be satisfied.

Inserting (1.3.4) and (1.3.5) into (1.3.3) we definitely obtain

$$\frac{d}{dt}\left[ N(t)\langle S\rangle(t)\right] = N(t)\langle Q\rangle(t). \tag{1.3.6}$$

If the number of particles is a constant or normalizing is performed for the total probability, then $N(t) = N_0 = const$, (or $N(t) = N_0 = 1$) and expression (1.3.6) is written as follows

$$\frac{d\langle S\rangle}{dt} = \langle Q\rangle. \tag{1.3.7}$$

The obtained equation (1.3.7) is an average analogue to initial equation (1.1.2) for the function $S$. Due to (1.3.2) the function $\langle S \rangle$ is associated with the entropy $H$, therefore for $H$ the following equation is true

$$\frac{d}{dt} H(t) = \frac{d}{dt}\langle S\rangle(t) = \langle Q\rangle,$$

$$\frac{dH}{dt} = \langle Q\rangle. \tag{1.3.8}$$

Equation (1.3.8) shows the change of the entropy in time. If the value $\langle Q\rangle(t)$ is positive, then the entropy increases; if it is negative, then the entropy decreases; and if it equals zero, then the entropy is a constant value. Taking into account expressions (1.1.7) and (1.1.31), we obtain the expression for $\langle Q\rangle(t)$

$$\langle Q\rangle(t) = \frac{\hbar n}{m} \int_0^{2\pi} d\phi \int_0^{\pi} r^2 \sin\theta d\theta \int_0^{+\infty} \frac{\text{ctg}\,\theta}{r^2} f(r,\phi,\theta,t) dr = \frac{\hbar n}{m} \int_0^{2\pi} d\phi \int_0^{\pi} \cos\theta d\theta \int_0^{+\infty} f(r,\phi,\theta,t) dr =$$

$$= \frac{\hbar n}{m} \int_0^{2\pi} d\phi \int_0^{\pi} \text{ctg}\,\theta d\theta \int_0^{+\infty} F_0(r,\phi,\theta) dr,$$

$$\langle Q\rangle(t) = \frac{\hbar n}{m} \int_0^{2\pi} d\phi \int_0^{\pi} \text{ctg}\,\theta d\theta \int_0^{+\infty} F_0\left( r, \phi + \frac{k}{n}\left( \text{ctg}\,\theta - \text{ctg}\left(\theta + \frac{2\alpha n}{r^2} t\right)\right), \theta + \frac{2\alpha n}{r^2} t\right) dr \tag{1.3.9}$$

Thus, the entropy $S(\vec{r},t)$ satisfies equation (1.1.2), and the entropy $H(t) = \langle S\rangle(t)$ satisfies equations (1.3.7), (1.3.8). The behavior of the entropies $S$ and $H$ with time is defined by the right sides $Q$ and $\langle Q\rangle$ of equations (1.1.2) and (1.3.8) respectively. The value of $Q$ is calculated (1.1.7), and its distribution is presented in Fig. 1. The value of $\langle Q\rangle$ may be calculated using formula (1.3.9).

In paragraph 2.3 an example is considered of the entropy calculation (1.3.9) for the model of «black» and «white» holes.

**§2. Examples**

This part represents examples of application of the mathematical model, $\Psi$-model, developed in paragraph 1 for micro and macro systems.

### 2.1 Motion in the modified Coulomb potential

Let us to consider the example when the agreement condition (1.1.55) don't take place. For example the Schrödinger equation solution might be given (1.1.45)

$$i\hbar\frac{\partial \Psi}{\partial t} = \frac{1}{2m}\left(\hat{p}-e\vec{A}\right)^2 \Psi + \tilde{U}\Psi, \tag{2.1.1}$$

with agree (1.1.11), (1.1.13) the vector potential $\vec{A}$ and magnetic field $\vec{B}$ have the form:

$$e\vec{A} = -\frac{\hbar}{\rho}\vec{e}_\phi, \ \vec{B} = \text{rot}\,\vec{A} = -\frac{q_m^{(Wb)}\delta(\rho)}{2\pi\rho}\vec{e}_z, \tag{2.1.2}$$

where $q_m^{(Wb)}q_e = 2\pi\hbar$, $\rho = r\sin\theta$. As the potential $\tilde{U}$ let us to take the modified Coulomb potential in the form (see Fig. 8)

$$\tilde{U}(r,\theta) = -\frac{Ze^2}{4\pi\varepsilon_0}\frac{1}{r} - \frac{\hbar^2\kappa}{2m}\frac{1}{r^2\sin^2\theta}, \tag{2.1.3}$$

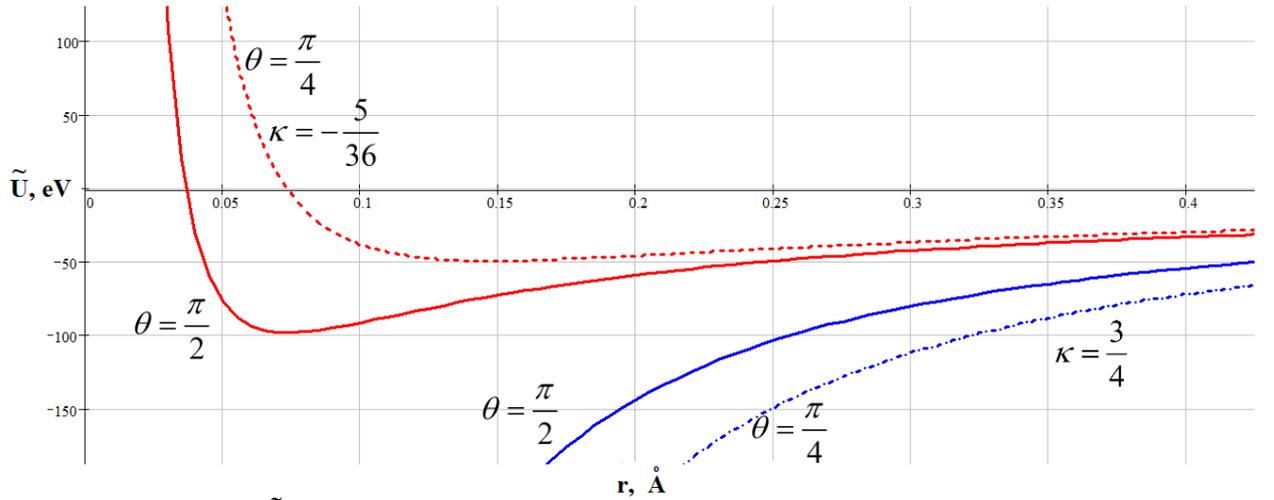

Fig. 8 Potential $\tilde{U}(r,\theta)$ distribution in case of $Z=1$

where $\kappa$ is constant value. The solution of the equation (2.1.1) with agree [27], (1.1.13), (1.1.39) let us will to find in the form

$$\Psi = |\Psi|e^{i\varphi} = \frac{R(r)}{\sqrt{\sin\theta}}e^{in\theta-i\frac{E}{\hbar}t} \overset{\text{det}}{=} \Psi_0(r,\theta)e^{-i\frac{E}{\hbar}t}. \tag{2.1.4}$$

The equation (2.1.1) with (2.1.3) and (2.1.4) may be written in the form

$$\left(\hat{p}-e\vec{A}\right)^2\Psi = \hat{p}^2\Psi - e(\hat{p},\vec{A}\Psi) - e(\vec{A},\hat{p}\Psi) + e^2|\vec{A}|^2\Psi = -\hbar^2\Delta\Psi + i\hbar e\Psi(\nabla,\vec{A}) +$$

$$+ i\hbar e(\nabla\Psi,\vec{A}) + i\hbar e(\vec{A},\nabla\Psi) + e^2|\vec{A}|^2\Psi,$$

because of $\operatorname{div}\vec{A}=0$ and $\vec{A}\perp\nabla\Psi$ (see (2.1.2), (2.1.4)) we have

$$\left(\hat{p}-e\vec{A}\right)^2\Psi=-\hbar^2\Delta\Psi+\frac{\hbar^2k^2}{r^2\sin^2\theta}\Psi,$$

$$\mathrm{E}\Psi=-\frac{\hbar^2}{2m}\Delta\Psi+\frac{\hbar^2k^2}{2mr^2\sin^2\theta}\Psi-\frac{Ze^2}{4\pi\varepsilon_0}\frac{1}{r}\Psi-\frac{\hbar^2\kappa}{2mr^2\sin^2\theta}\Psi,$$

or taking into account (2.1.4)

$$\frac{\Delta\Psi_0}{\Psi_0}=-\frac{2m\mathrm{E}}{\hbar^2}+\frac{k^2-\kappa}{r^2\sin^2\theta}-\frac{2m}{\hbar^2}\frac{Ze^2}{4\pi\varepsilon_0}\frac{1}{r}. \qquad (2.1.5)$$

Let us calculate $\dfrac{\Delta\Psi_0}{\Psi_0}$

$$\Delta\Psi_0=\frac{e^{in\theta}}{\sqrt{\sin\theta}}\Delta_r R+\frac{R}{r^2\sin\theta}\frac{\partial}{\partial\theta}\sin\theta\frac{\partial}{\partial\theta}\frac{e^{in\theta}}{\sqrt{\sin\theta}},$$

$$\frac{\Delta\Psi_0}{\Psi_0}=\frac{\Delta_r R}{R}+\frac{1}{r^2}\left(\frac{1+\sin^2\theta}{4\sin^2\theta}-n^2\right). \qquad (2.1.6)$$

Substituting (2.1.6) into (2.1.5), will be given

$$\frac{\Delta_r R}{R}+\frac{1}{r^2}\left(\frac{1+\sin^2\theta}{4\sin^2\theta}-n^2\right)=-\frac{2m\mathrm{E}}{\hbar^2}+\frac{k^2-\kappa}{r^2\sin^2\theta}-\frac{2m}{\hbar^2}\frac{Ze^2}{4\pi\varepsilon_0}\frac{1}{r},$$

$$R''+\frac{2}{r}R'+\left(\frac{2m\mathrm{E}}{\hbar^2}-\frac{4(k^2-\kappa)-1-\sin^2\theta}{4r^2\sin^2\theta}-\frac{n^2}{r^2}+\frac{2m}{\hbar^2}\frac{Ze^2}{4\pi\varepsilon_0}\frac{1}{r}\right)R=0. \qquad (2.1.7)$$

Let us select the parameter $k$ as $4(k^2-\kappa)-1=0$, that is

$$k^2=\kappa+\frac{1}{4}, \qquad (2.1.8)$$

then equation (2.1.7) will be written in the form

$$R''+\frac{2}{r}R'+\left(\frac{2m\mathrm{E}}{\hbar^2}-\frac{n^2-\frac{1}{4}}{r^2}+\frac{2m}{\hbar^2}\frac{Ze^2}{4\pi\varepsilon_0}\frac{1}{r}\right)R=0. \qquad (2.1.9)$$

Let us change the variable $s=\lambda r$, $R(r)\stackrel{\text{det}}{=}R_0(s)$, where $\lambda=\dfrac{2}{\hbar}\sqrt{-2m\mathrm{E}}$, then the equation (2.1.9) may be written in the form

$$R_0'' + \frac{2}{s}R_0' + \left(-\frac{1}{4} + \frac{\nu}{s} - \frac{l(l+1)}{s^2}\right)R_0 = 0, \qquad (2.1.10)$$

where

$$\nu \stackrel{det}{=} \frac{2m}{\hbar^2 \lambda} \frac{Ze^2}{4\pi\varepsilon_0} = \frac{1}{\hbar}\sqrt{\frac{m}{-2E}} \frac{Ze^2}{4\pi\varepsilon_0}, \quad n^2 - \frac{1}{4} \stackrel{det}{=} l(l+1),$$

or

$$E_\nu = -\frac{Z^2 e^4 m}{32\pi^2 \varepsilon_0^2 \hbar^2 \nu^2}, \quad n = l + \frac{1}{2}. \qquad (2.1.11)$$

The equation solution (2.1.10) may be written by the hypergeometric function F

$$R_0(s) = s^l e^{-s/2} F(-\nu + l + 1, 2l + 2, s), \qquad (2.1.12)$$

where because of the normalization condition the value $-\nu + l + 1$ is negative and integer one, or is equal to zero one. That is $\nu \geq l+1$. Taking into account (2.1.12) and (2.1.4) the equation solution (2.1.1) obtained

$$\Psi_{l\nu}(r,\theta) = C_0 \frac{e^{i\left(l+\frac{1}{2}\right)\theta}}{\sqrt{\sin\theta}} (r\lambda_\nu)^l e^{-r\lambda_\nu/2} F(-\nu + l + 1, 2l + 2, r\lambda_\nu) e^{-i\frac{E_\nu}{\hbar}t}, \qquad (2.1.13)$$

where $\lambda_\nu = \frac{2}{\hbar}\sqrt{-2mE_\nu}$, $C_0$ is the constant value, which from normalization condition will be choose.

As a consequence of the violation of condition (1.1.55) in the form of potential $\sim \frac{1}{r}$ (2.1.3) instead of $\sim \frac{1}{r^2}$ for the vector field $\langle\vec{v}\rangle$ (1.1.13) it leads to the existence of the discrete set of trajectories with the discrete set of energies instead of the trajectory continuum (2.1.8), (2.1.11), which was to be proved.

*Remark 5.*

*For the Coulomb potential $\sim \frac{1}{r}$, one can choose a velocity vector field $\langle\vec{v}\rangle$ different from (1.1.13), for the self-consistency conditions (1.1.55) not to be violated. For example, the velocity field may be chosen $|\langle\vec{v}\rangle| \sim \frac{1}{\sqrt{r}}$, then kinetic and potential energy have the same order. For example, the potential of gravitational interaction has the order $\sim \frac{1}{r}$ as well as the Coulomb potential. The field of circular rotation velocities of a certain point object about the Earth has dependencies of the form $\langle\vec{v}\rangle = \sqrt{\frac{MG}{r}}\vec{e}_\phi$, where M is the Earth mass, and G is the gravitational constant. As a result, the conditions of the self-consistency of kinetic and potential energy will be met.*

### 2.2 Motion in a self-consistency field

Let us consider a steady-state case of equation (1.1.14)

$$\frac{k}{\sin^2\theta}\frac{\partial S}{\partial \phi}+n\frac{\partial S}{\partial \theta}=n\,\text{ctg}\,\theta. \qquad (2.2.1)$$

The solution of equation (2.2.1), according to (1.1.20)-(1.1.22) and (1.1.30), is of the following form

$$S_{g.het.}(r,\phi,\theta)=G_0\left(r,\phi+\frac{k}{n}\text{ctg}\,\theta\right)+\ln\sin\theta. \qquad (2.2.1)$$

By analogy with re-expressions (1.1.31), (1.1.32), we obtain for the probability density function

$$f(r,\phi,\theta)=\frac{1}{\sin\theta}F_0\left(r,\phi+\frac{k}{n}\text{ctg}\,\theta\right), \qquad (2.2.2)$$

where the function $F_0$ is defined from the boundary conditions, e.g., at $\theta=\frac{\pi}{2}$

$$f(r,\phi,\theta)\big|_{\theta=\frac{\pi}{2}}=F_0\left(r,\phi,\frac{\pi}{2}\right)=f_0(r,\phi). \qquad (2.2.3)$$

The function $f_0(r,\phi)$ determines the probability density distribution in the median plane XOY. Taking into account (2.2.3), solution (2.2.2) is as follows

$$f(r,\phi,\theta)=\frac{1}{\sin\theta}f_0\left(r,\phi+\frac{k}{n}\text{ctg}\,\theta\right). \qquad (2.2.4)$$

Let the boundary function $f_0(r,\phi)$ satisfy

$$f_0(r,\phi)=R^2(r)Y^2(\phi). \qquad (2.2.5)$$

In this case solution (2.2.4) is as follows

$$f(r,\theta,\phi)=\frac{1}{\sin\theta}R^2(r)Y^2\left(\phi+\frac{k}{n}\text{ctg}\,\theta\right). \qquad (2.2.6)$$

For the wave function $\Psi$ due to (1.1.39), (2.2.6), the following expression is true

$$\Psi(\vec{r})=\frac{R(r)}{\sqrt{\sin\theta}}Y\left(\phi+\frac{k}{n}\text{ctg}\,\theta\right)e^{in\theta}. \qquad (2.2.6)$$

Let us find the corresponding potential $U$ (1.1.40) for the wave function (2.2.6). Let us calculate $\Delta|\Psi|$. Let us designate for computational convenience

$$\Lambda(\theta,\phi) \stackrel{\text{det}}{=} \frac{1}{\sqrt{\sin\theta}} Y\left(\phi + \frac{k}{n}\operatorname{ctg}\theta\right), \qquad (2.2.7)$$

we obtain

$$\frac{\Delta|\Psi|}{|\Psi|} = \frac{\Delta[R(r)\Lambda(\theta,\phi)]}{R(r)\Lambda(\theta,\phi)} = \frac{\Lambda\Delta_r R + \dfrac{R}{r^2}\Delta_{\theta,\phi}\Lambda}{R\Lambda} = \frac{\Delta_r R}{R} + \frac{1}{r^2}\frac{\Delta_{\theta,\phi}\Lambda}{\Lambda}, \qquad (2.2.8)$$

Inserting (2.2.8) into (1.1.40), we obtain

$$U(\vec{r}) = \frac{\alpha}{\beta}\left[\frac{1}{r^2}\left(n^2 - \frac{\Delta_{\theta,\phi}\Lambda}{\Lambda}\right) - \frac{\Delta_r R}{R}\right]. \qquad (2.2.9)$$

The expression for $\dfrac{\Delta_r R}{R}$ is as follows

$$\frac{\Delta_r R}{R} = \frac{R''}{R} + \frac{2}{r}\frac{R'}{R}. \qquad (2.2.10)$$

Let us rewrite the expression for $\Delta_{\theta,\phi}\Lambda$

$$\Delta_{\theta,\phi}\Lambda = \frac{1}{\sin\theta}\frac{\partial}{\partial\theta}\sin\theta\frac{\partial}{\partial\theta}\frac{Y}{\sqrt{\sin\theta}} + \frac{1}{\sin^2\theta\sqrt{\sin\theta}}\frac{\partial^2 Y}{\partial\phi^2} =$$

$$= \frac{1}{\sin\theta}\frac{\partial}{\partial\theta}\left(Y'_\theta\sqrt{\sin\theta} - Y\frac{\cos\theta}{2\sqrt{\sin\theta}}\right) + \frac{1}{\sin^{5/2}\theta}Y''_{\phi\phi} =$$

$$= \frac{1}{\sqrt{\sin\theta}}Y''_{\theta\theta} + Y'_\theta\frac{\cos\theta}{2\sin^{3/2}\theta} - Y'_\theta\frac{\cos\theta}{2\sin^{3/2}\theta} - Y\frac{-\sin^{3/2}\theta - \dfrac{\cos^2\theta}{2\sqrt{\sin\theta}}}{2\sin^2\theta} + \frac{1}{\sin^{5/2}\theta}Y''_{\phi\phi} =$$

$$= \frac{1}{\sqrt{\sin\theta}}Y''_{\theta\theta} + Y\frac{2\sin^2\theta + \cos^2\theta}{4\sin^{5/2}\theta} + \frac{1}{\sin^{5/2}\theta}Y''_{\phi\phi} = \frac{1}{\sqrt{\sin\theta}}\left(Y''_{\theta\theta} + Y\frac{1+\sin^2\theta}{4\sin^2\theta} + \frac{1}{\sin^2\theta}Y''_{\phi\phi}\right).$$

The result is as follows

$$\frac{\Delta_{\theta,\phi}\Lambda}{\Lambda} = \frac{Y''_{\theta\theta}}{Y} + \frac{1}{\sin^2\theta}\frac{Y''_{\phi\phi}}{Y} + \frac{1+\sin^2\theta}{4\sin^2\theta}. \qquad (2.2.11)$$

As $Y(\xi)$ is a function of single variable, then it is needed to transform the partial derivatives $Y''_{\phi\phi}$ and $Y''_{\theta\theta}$ in expression (2.2.11). Taking into account expression (1.1.23), expression (2.2.11) is as follows

$$Y''_{\phi\phi} = Y'', \quad Y'_{\theta} = -\frac{k}{n\sin^2\theta}Y', \quad Y''_{\theta\theta} = \frac{k^2}{n^2\sin^4\theta}Y'' + 2\frac{k\operatorname{ctg}\theta}{n\sin^2\theta}Y',$$

$$\frac{\Delta_{\theta,\phi}\Lambda}{\Lambda} = \frac{Y''}{Y}\frac{k^2}{n^2\sin^4\theta} + 2\frac{k\operatorname{ctg}\theta}{n\sin^2\theta}\frac{Y'}{Y} + \frac{1}{\sin^2\theta}\frac{Y''}{Y} + \frac{1+\sin^2\theta}{4\sin^2\theta},$$

$$\frac{\Delta_{\theta,\phi}\Lambda}{\Lambda} = \left(1 + \frac{k^2}{n^2\sin^2\theta}\right)\frac{Y''}{Y\sin^2\theta} + 2\frac{k\operatorname{ctg}\theta}{n\sin^2\theta}\frac{Y'}{Y} + \frac{1+\sin^2\theta}{4\sin^2\theta}. \qquad (2.2.12)$$

The expressions (2.2.10), (2.2.12) are needed to find the potential (2.2.9). Let us consider as an example boundary condition (2.2.5) of the following form

$$f_0(r,\phi) = C^2 r^{2\upsilon} e^{-2\kappa r} \sin^4(l\phi),$$
$$R(r) = Cr^{\upsilon} e^{-\kappa r}, \quad Y(\phi) = \sin^2(l\phi). \qquad (2.2.13)$$

Solution (2.2.6) is as follows

$$f(r,\theta,\phi) = \frac{C^2 r^{2\upsilon} e^{-2\kappa r}}{\sin\theta} \sin^4\left[l\left(\phi + \frac{k}{n}\operatorname{ctg}\theta\right)\right], \qquad (2.2.14)$$

To be definite, let us take $k=1$ and $n=3$. The probability of finding in the volume of space $d\omega = r^2 \sin\theta d\theta d\phi dr$ with the center in point $(r,\theta,\phi)$ is $dP = f(r,\theta,\phi)d\omega$. In Figs. 9-11 the angular probability distribution $dP$ is shown on the spherical layer of the fixed radius at $l = 1,2,3$. The surface shown in Figs. 9-11 is described by a radius-vector:

$$\vec{r} = \{x^1, x^2, x^3\}, \qquad (2.2.15)$$
$$x^1 = dP\sin\theta\cos\phi, \quad x^2 = dP\sin\theta\sin\phi, \quad x^3 = dP\cos\theta$$

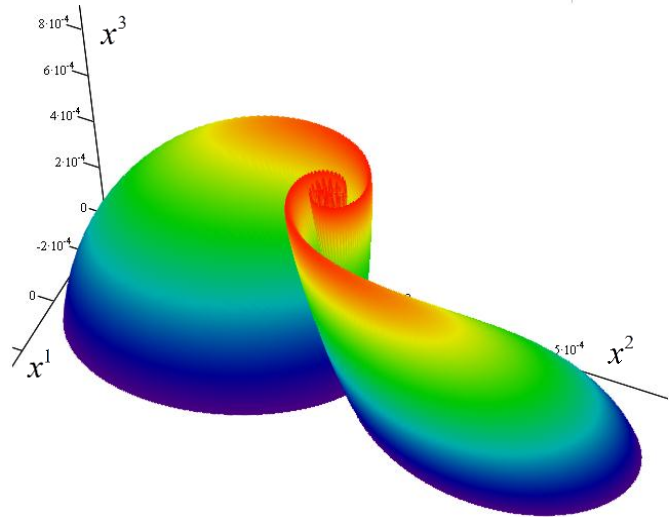

Fig. 9 Angular distribution of probability at $l = 1$

As boundary condition (2.2.13) (at $\theta = \frac{\pi}{2}$) determines the periodic angular probability distribution $dP$, according to the function $\sin^4(l\phi)$, then the maximum probability is achieved at

«tops», which number is determined as $2l$. The spiral structure of the chains (at $0 < \theta < \frac{\pi}{2}$) is caused by the spiral structure of characteristics (1.1.23) (see Fig. 5).

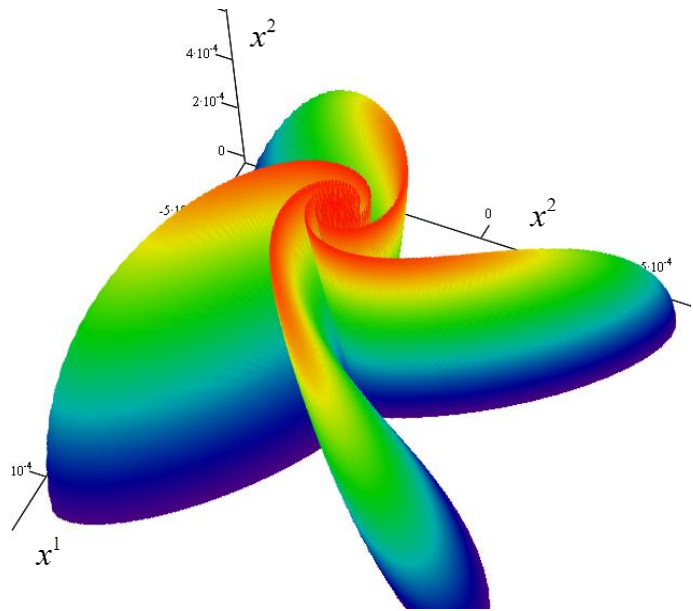

Fig. 10 Angular distribution of probability at $l = 2$

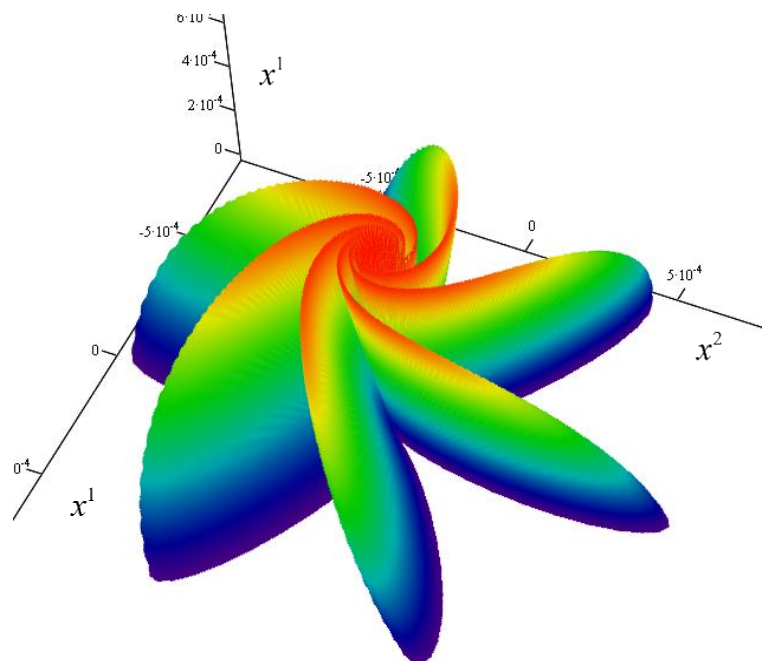

Fig. 11 Angular distribution of probability at $l = 3$

Let us find potential $U$ corresponding to wave function (2.2.6) at given boundary conditions (2.2.13). According to expressions (2.2.10) and (2.2.12), we obtain

$$\frac{\Delta_r R}{R} = \kappa^2 - \frac{2\kappa(\upsilon+1)}{r} + \frac{\upsilon(\upsilon+1)}{r^2} \qquad (2.2.16)$$

$$\frac{\Delta_{\theta,\phi}\Lambda}{\Lambda} = \left(1 + \frac{k^2}{n^2\sin^2\theta}\right)\left(\frac{1}{\sin^2(m\xi)} - 2\right)\frac{2m^2}{\sin^2\theta} + 4m\frac{k\,\text{ctg}\,\theta}{n\sin^2\theta}\text{ctg}(m\xi) + \frac{1+\sin^2\theta}{4\sin^2\theta}$$

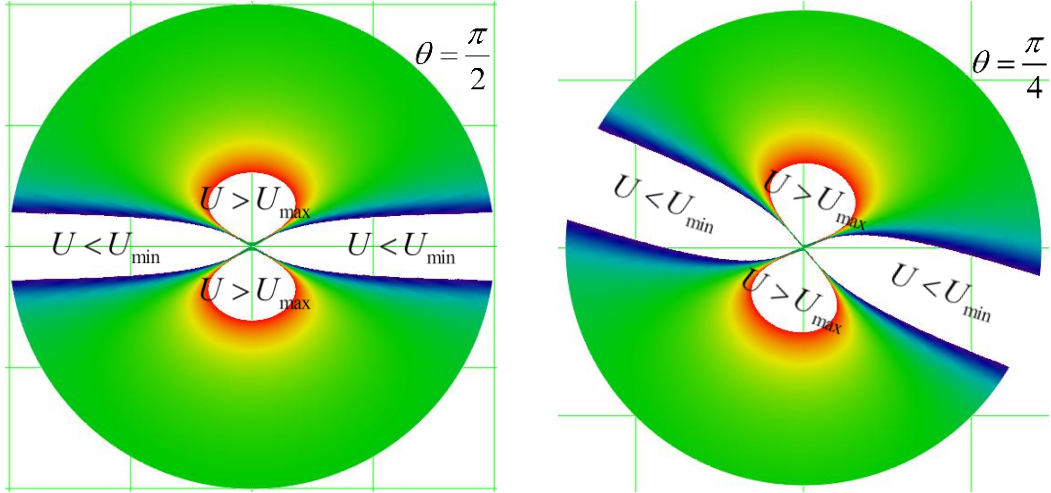

Fig. 12 Potential $U(r,\phi)$ distribution in case of $l=1$

Inserting (2.2.16) into (2.2.9), let us build the distribution of потенциала $U(r,\theta,\phi)$ in the horizontal layers (parallel to the plane XOY at $\theta = \frac{\pi}{2}, \frac{\pi}{4}$) for the variants $l=1,2,3$. From expression (2.2.16) it follows that the value of potential $U(r,\theta,\phi)$ may tend to infinity, at $\sin^2(m\xi) \to 0$, such regions correspond to the potential wells. As the value of the potential may change the sign, then there will be potential barriers besides potential wells.

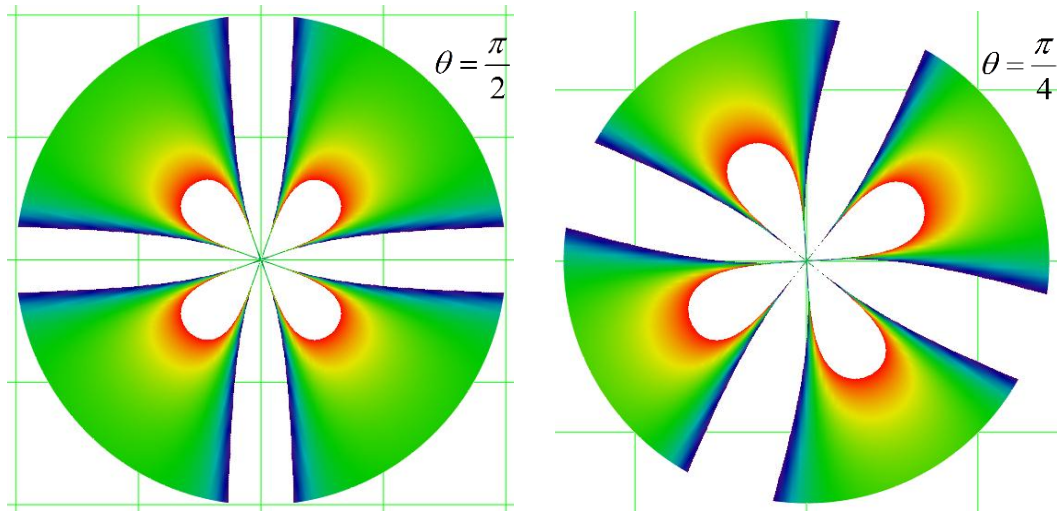

Fig. 13 Distribution of potential $U(r,\phi)$ in case of $l=2$

Due to the presence of the potential wells and potential barriers, for visual reference, we «cut» the value of the potentials with the maximum $U_{max}$ and minimum $U_{min}$ value. Therefore the value $U < U_{min}$ (potential well) and $U > U_{max}$ (potential barrier) of the potential are not shown. The empty space in Figure corresponds to the regions of the potential wells and barriers.

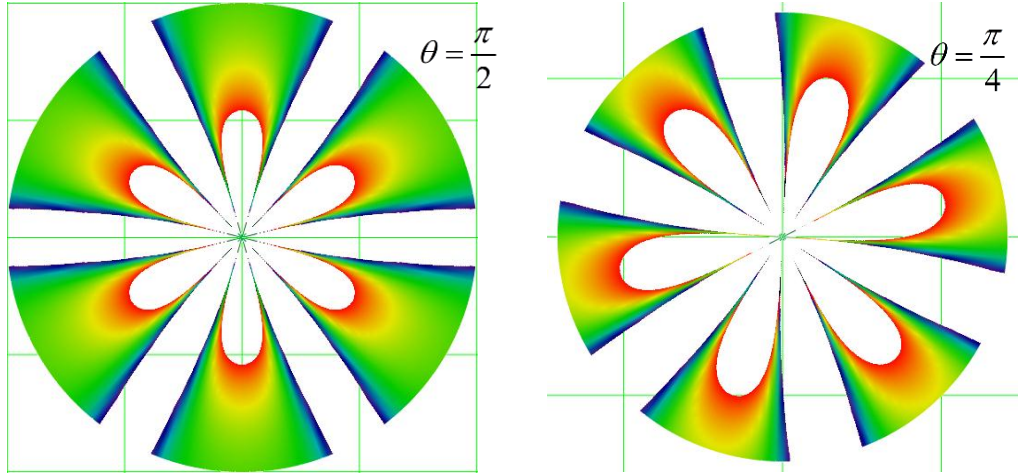

Fig. 14 Distribution of potential $U(r,\phi)$ in case of $l = 3$

In Figs. 12-14 the distributions of potentials $U(r,\phi)$ are shown corresponding to $l = 1,2,3$. Each of the mentioned figures shows two distributions $U(r,\phi)$, which correspond $\theta = \frac{\pi}{2}$ (on the left) and $\theta = \frac{\pi}{4}$ (on the right). In Fig. 12 the regions of the potential wells and barriers are explicitly indicated, therefore, in order not to overload the image, Figs. 13-14 do not contain analog designations. Figs.12-14 show that the potential barriers separate the potential wells from each other.

The comparison of the probability distributions $dP$ (see Figs. 9-11) with the distribution of potentials $U(r,\phi)$ (see Fig. 12-14) corresponding to the same value of parameter $l$ shows that the regions of the maximum probability (tops) coincide with the regions of the potential wells.

As a result, the maximum probability $dP$ is focused in the spiral potential wells, and its flow $\langle \vec{v} \rangle$ (1.1.6) due to (1.1.36) is directed at the tangent to characteristics (1.1.23) and flowing through one pole of the sphere to another (see Figs. 3-5).

## 2.3 Astrophysics model

As the mathematical model ($\Psi$-model) obtained in paragraph 1 has the general result not only for quantum mechanics but also for hydro- and aerodynamics, then the probability density function $f$ may be interpreted as a function of mass density or charge one and the probability flow velocity $\langle \vec{v} \rangle$ as a function of flow rate of fluid, gas or charge.

Application of the developed $\Psi$-model to astrophysics seems to be feasible. The spiral structure of the galaxy has the arms converging in its center. Theoretically, a black (white) hole may be the center of the spiral galaxy (anti-galaxy). White hole have not been observed yet. However, the full Schwarzschild solution contains both the black and white holes [33].

The diagrams of the characteristic (trajectories being the solution of motion equation (1.1.47)) presented in Figs. 3-5,15 have spiral structure as well as spiral galaxy arms. At that, the velocity vector $\langle\vec{v}\rangle$ is directed at the tangent to the spiral trajectories (1.1.36). The spiral trajectories are directed from one pole of the sphere to the opposite one and tend to the poles infinitely long time (see Remark 4, Fig. 8).

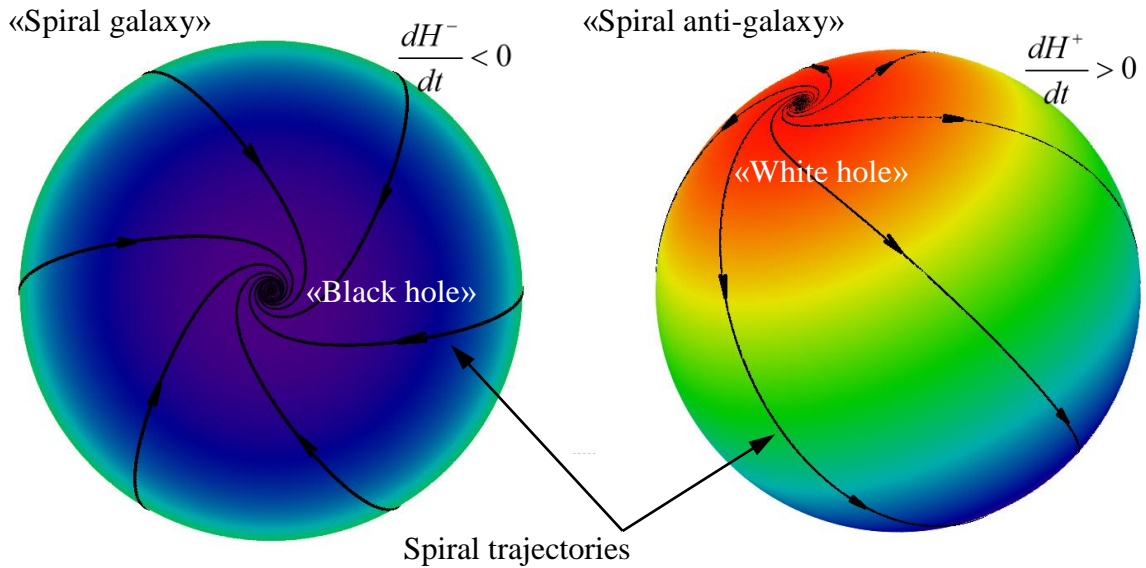

Fig. 15 Astrophysics analogy

At the sphere poles ($\theta = 0$ and $\theta = \pi$), the density $f$ tends to infinity (1.1.31), which may correspond to massive holes (see Fig. 3-5,15). Thus, the pole, where the spiral trajectories arrive, may be interpreted as a «black hole», and the pole, from which the spiral trajectories emerge, as a «white hole» (see Fig.15).

On the axis connecting the opposite poles («black» and «white hole») the velocity vector field is not define, as the vector rotates about the OZ axis with the infinite velocity and does not have particular direction, and the velocity modulus $|\langle\vec{v}\rangle|$ tends to infinity. The intrinsic moment $\vec{\mu}_s$ of the system is directed along the OZ axis.

«In 1968 Carter noticed that the Kerr-Newman (KN) solution has gyromagnetic ratio g = 2, as that of the Dirac electron, and therefore, at least the asymptotic electromagnetic and gravitational fields of the electron should correspond to the KN solution with great precision» [44-47]. The gyromagnetic ratio g in $\Psi$ - model is equal 2 (1.2.7) so.

The axis connecting the opposite poles may be interpreted as a wormhole. Fig. 16 presents the geometric interpretation of the wormhole with the general relativity theory standpoint [34]. As is known, wormholes fall into intra-universe and inter-universe, depending on whether it is possible to connect their mouths via a curve not crossing the throat (see Fig. 16). The comparison between Figs. 15 and 16 shows that there are two ways to connect the sphere (hole) poles: along the axis and along the sphere surface.

As it is known the black hole rotation leads to event horizon radius decreasing (Kerr solution). This effect is known as the «naked singularity» in general relativity [35-39].

The rotation velocity at poles (on the OZ axis) tends to infinity in $\Psi$ - model. Propose event horizon radius tends to zero. The mass (charge) density tends to infinity (1.1.31). In this case OZ axis appears as event horizon (see Fig. 15, 21).

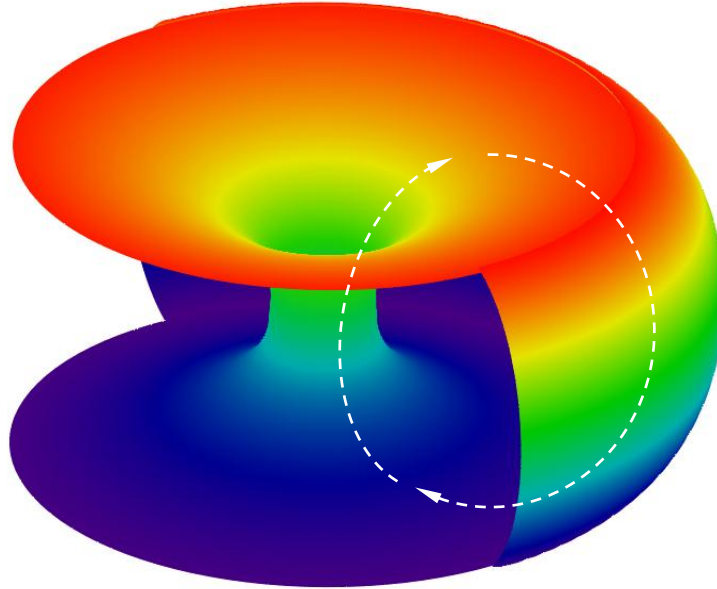

Рис. 16 Two-dimensional section of a simple wormhole

The source of the velocity field $\langle \vec{v} \rangle$ is value $Q$ (1.1.7) (see Fig. 1), which tends to infinity at an axis connecting «black and white holes» (poles of the sphere). The top part of the full-sphere has positive sources $Q > 0$, and the bottom one has negative sources $Q < 0$. Thus, the bottom part of the full-sphere corresponds to the spiral galaxy absorbing mass ($Q < 0$), and the top part corresponds to the «anti-galaxy» throwing mass out ($Q > 0$).

Let us consider the entropy behavior in the neighborhood of the sphere poles («black» and «white» hole). According to equations (1.1.2) and (1.3.7) the behavior of the entropies $S$ and $H$ is determined by $Q$ and $\langle Q \rangle$ respectively. The value $Q$ is known (1.1.7), therefore let us calculate $\langle Q \rangle$ (1.3.9). As $\langle Q \rangle$ depends on the initial (at $t = 0$) density of distribution $F_0$ (1.3.8), we take as an example

$$F_0(r, \phi, \theta) = R^2(r)\sin^2(l\phi),\ l \in \mathbb{N} \qquad (2.3.1)$$

Inserting (2.3.1) into (1.3.9), we obtain

$$\langle Q \rangle(t) = \frac{\hbar n}{m} \int_0^{2\pi} d\phi \int_0^{\pi} \operatorname{ctg}\theta\, d\theta \int_0^{+\infty} R^2(r)\sin^2 l\left(\phi + \bar{\omega}_{k,n}\right) dr, \qquad (2.3.2)$$

Where we designate

$$\bar{\omega}_{k,n}(r, \theta) \stackrel{\text{det}}{=} \frac{k}{n}\left(\operatorname{ctg}\theta - \operatorname{ctg}\left(\theta + \omega_n(r)t\right)\right). \qquad (2.3.3)$$

As $\bar{\omega}_{k,n}(r, \theta)$ does not depend on $\phi$, integral (2.3.2) may be rewritten as follows

$$\langle Q \rangle(t) = \frac{\hbar n}{m} \int_0^{2\pi} \sin^2 l(\phi + \bar{\omega}_{k,n}) d\phi \int_0^{\pi} \operatorname{ctg} \theta d\theta \int_0^{+\infty} R^2(r) dr =$$

$$= \frac{\hbar n}{2m} \int_0^{2\pi} \left(1 - \cos\left[2l(\phi + \bar{\omega}_{k,n})\right]\right) d\phi \int_0^{\pi} \operatorname{ctg} \theta d\theta \int_0^{+\infty} R^2(r) dr = \frac{\pi \hbar n}{m} \int_0^{\pi} \operatorname{ctg} \theta d\theta \int_0^{+\infty} R^2(r) dr,$$

that is

$$\langle Q \rangle(t) = 0. \tag{2.3.4}$$

Consequently, according to equation (1.3.7), the relative entropy $H$ is constant through the entire space

$$H = const. \tag{2.3.5}$$

Note that at averaging in integral (1.3.9) only over the top semisphere ($\varepsilon \leq \theta \leq \frac{\pi}{2}$).

$$\langle Q \rangle^+ \stackrel{\text{det}}{=} \frac{\pi \hbar n}{m} \int_\varepsilon^{\pi/2} \operatorname{ctg} \theta d\theta \int_0^{+\infty} R^2(r) dr, \tag{2.3.6}$$

where $\varepsilon$ – a certain small value, $\langle Q \rangle^+ > 0$ the entropy $H^+$ grows (see Fig.15). At $\varepsilon \to 0$ the value of $\langle Q \rangle^+ \to +\infty$ as the magnitude of $\theta = 0$ corresponds to the OZ axis connecting the poles (wormhole) of the sphere («black» and «white» holes). The region of the entropy growth corresponds to the exterior of the «white» hole (see Fig. 15).

At averaging in integral (1.3.9) over the bottom semisphere ($\frac{\pi}{2} \leq \theta \leq \pi - \varepsilon$)

$$\langle Q \rangle^- \stackrel{\text{det}}{=} \frac{\pi \hbar n}{m} \int_{\pi/2}^{\pi-\varepsilon} \operatorname{ctg} \theta d\theta \int_0^{+\infty} R^2(r) dr, \tag{2.3.7}$$

we obtain the magnitude of $\langle Q \rangle^- < 0$ and the entropy $H^-$ goes down (see Fig. 15). At $\varepsilon \to 0$ the value $\langle Q \rangle^- \to -\infty$ as the magnitude $\theta = \pi$ corresponds to the OZ axis, connecting the poles (wormhole) of the sphere («black» and «white» holes). The region of the entropy decrease corresponds to the exterior of the «black» hole (see Fig. 15).

If the initial density distribution (2.3.1) is symmetric in $\theta$ the terms corresponding the top semisphere $\langle Q \rangle^+$ (2.3.6) and the bottom one $\langle Q \rangle^-$ (2.3.7) cancel out each other and we obtain $\langle Q \rangle = 0$ (2.4.4). As the result of the total averaging over the entire space, according to (2.3.4), gives $\langle Q \rangle = 0$ and the entropy $H = const$ (2.3.5).

Figs. 17-20 show probability distribution (2.2.15) for unsteady solution (1.1.31) in case of initial density distribution (2.3.1). Remark 3 shows that solution (1.1.31) is periodic on each concentric sphere of radius $r$ with period $T$ and frequency $\omega_n(r)$

$$T(r) = \frac{m}{\hbar n}\pi r^2 = \frac{m}{\hbar n}\frac{\sigma}{4}, \qquad \omega_n(r) = \frac{\hbar n}{mr^2}, \qquad (2.3.8)$$

where $\sigma = 4\pi r^2$ is surface area of the sphere of radius $r$. Figs. 17-20 show the distributions corresponding to various time moments in period $T_0 = T(r_0)$, that is $t = 0, \frac{T_0}{4}, \frac{T_0}{2}, \frac{3T_0}{4}$. As the process of the density redistribution is periodic, then the distribution in the instant of time $t = 0$ (see Fig. 17) coincides with the instant of time $t = T_0$.

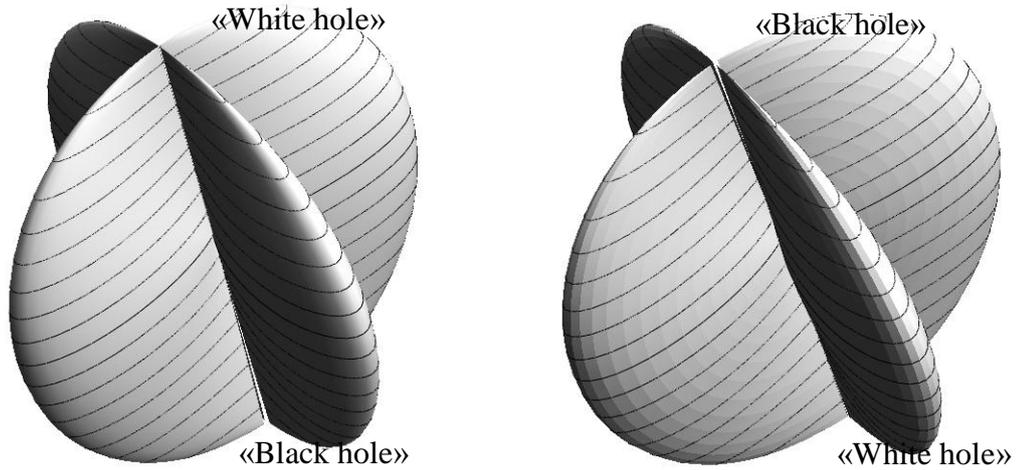

Fig. 17 Probability distribution on the sphere surface in the instant $t = 0$

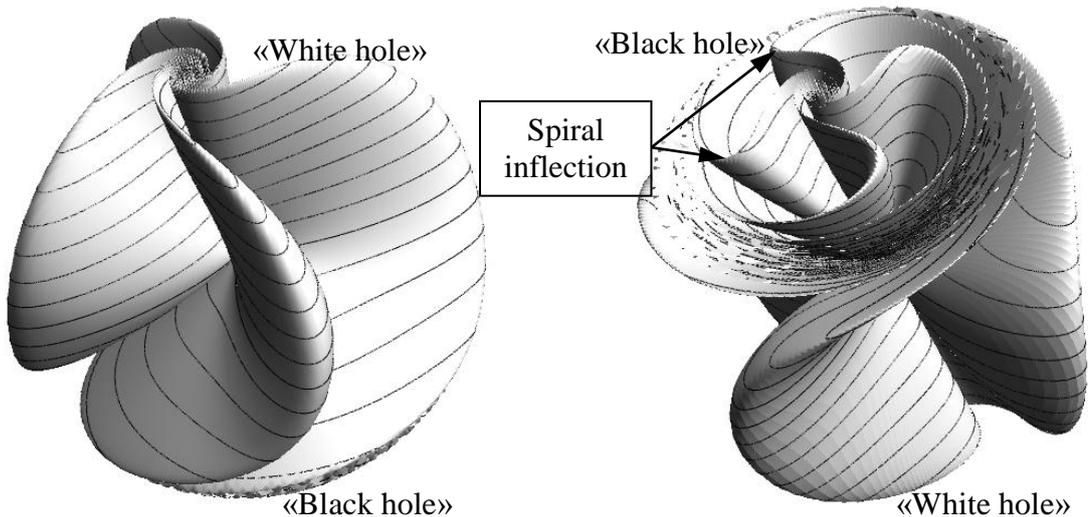

Fig. 18 Probability distribution on the sphere surface in the instant $t = \frac{T_0}{4}$

Each Figure from 17 to 20 illustrates two aspect views of the same distribution, but looking from the opposite poles («black» and «white» holes). The left figure corresponds to the view from the semisphere with $Q > 0$ («white» hole, «anti-galaxy»), the right figure corresponds to the view from the semisphere with $Q < 0$ («black» hole, «galaxy»).

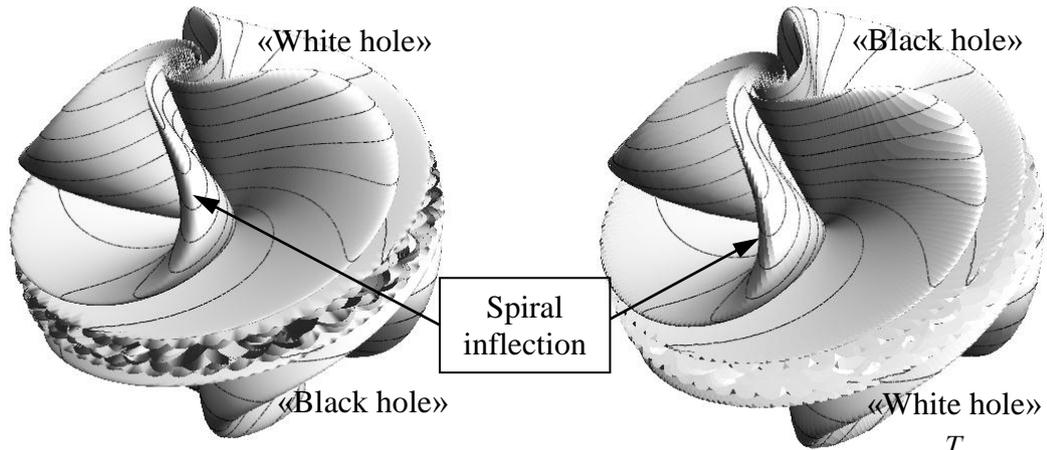

Fig. 19 Probability distribution on the sphere surface in the instant $t = \dfrac{T_0}{2}$

Comparison of Figs. 17-20 demonstrates that the distributions are generally not symmetric, except for the instants of time $t = 0, \dfrac{T_0}{2}, T_0$, still the entropy $H$ remains constant through the entire period according to (2.3.5).

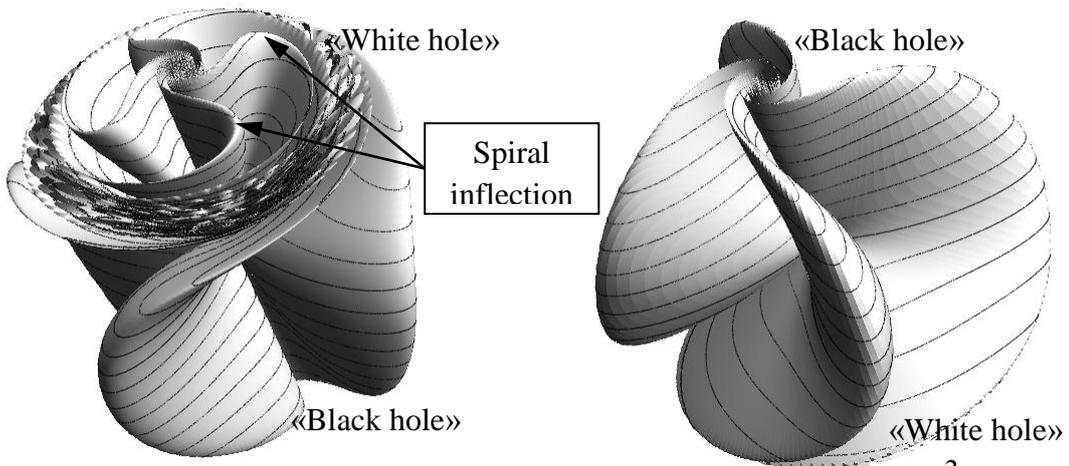

Fig. 20 Probability distribution on the sphere surface in the instant $t = \dfrac{3}{4}T_0$

Fig. 21 show the probability distributions on the ROZ cross-section corresponding to various time moments $t = T_0, 5T_0, 10T_0, 20T_0$.

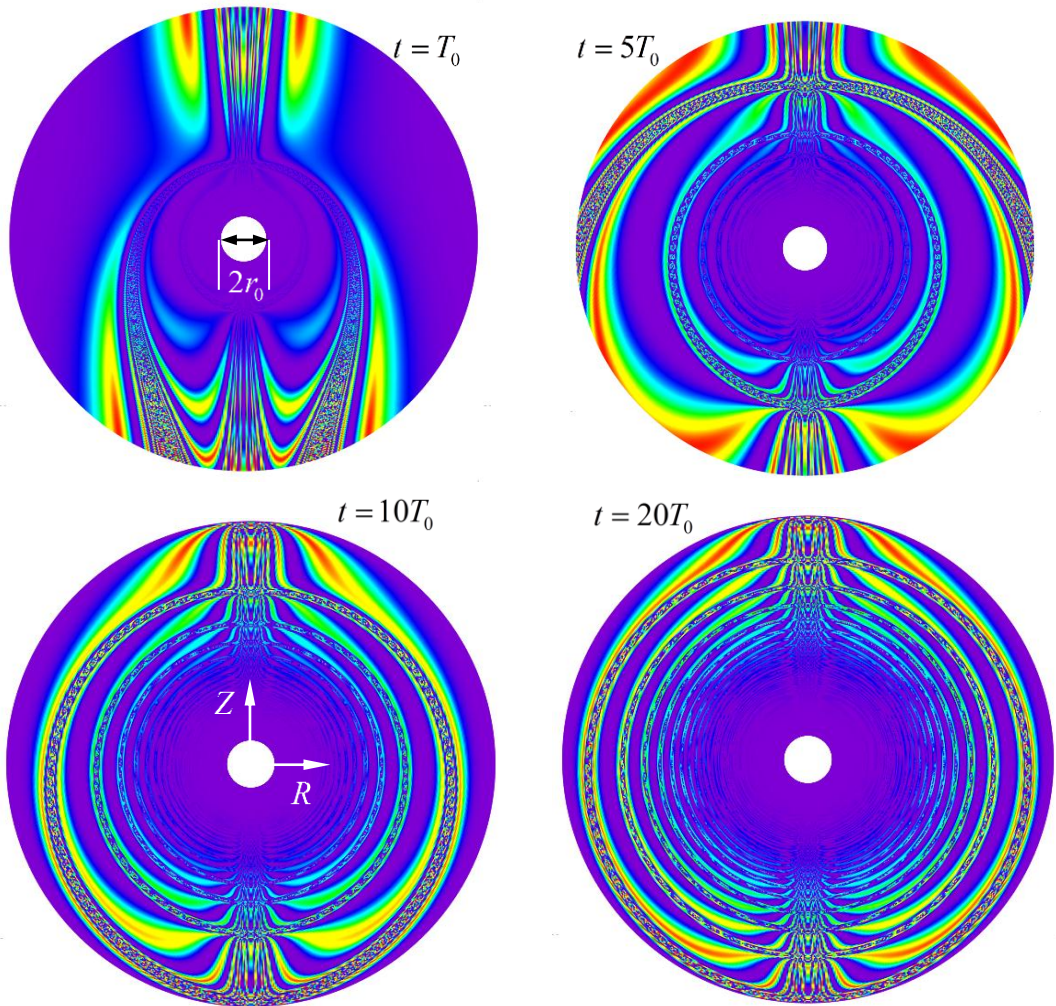

Fig. 21 Probability distribution on the ROZ cross-section

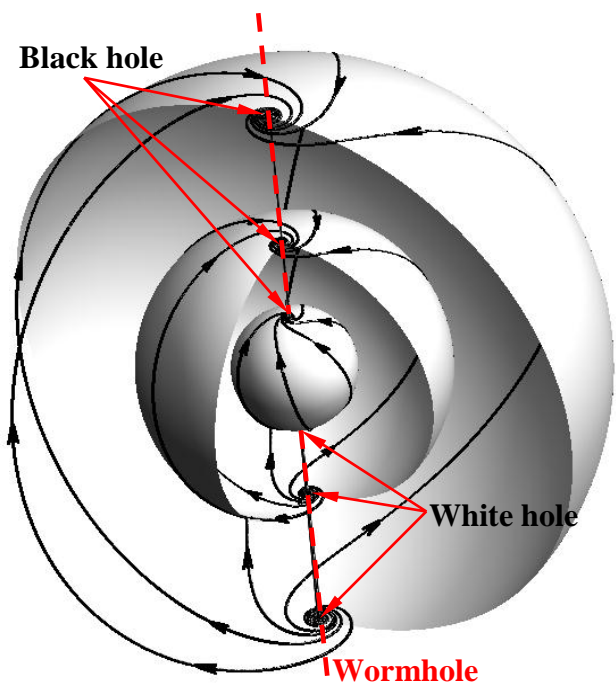

Fig. 22 Model of «enclosed worlds»

The growth and decrease of the entropies $H^+$ (2.4.6) and $H^-$ (2.4.7) are unaffected in the neighborhood of the «white» and «black» holes respectively.

Spiral structure of density distributions starts simultaneously at both poles of the sphere but with the different speed. It is seen from Figs. 17-20 that the spiral structure of the density distributions goes faster in the beginning (see Fig. 18) at the pole corresponding to the «black» hole. Gradually the created vortex moves to the opposite pole corresponding to the «white» hole (see Figs. 19-20). In the instant $t = \dfrac{T}{2}$ the speeds of whirling become equal at both poles (see Fig. 19). At $\frac{T}{2} < t < T$ the whirling process in the region of the «black» hole slows down, and the whirling process in the region of the «white» hole gains pace.

The mentioned inequality of the whirling speed of the poles leads to emergence of inflection of spiral density distributions (see Figs. 11-14).

The density redistribution happens periodically on each concentric sphere, but with different periods (2.3.8) (see Fig. 21).

Note that the described analogs with the spiral trajectories, poles are true for all concentric spheres with the common OZ axis (see Fig. 22).

Kerr-Newman metrics (Kerr, Reissner-Nordström) may be analytically continued through the event horizon by those way that set independent spaces in the black hole to connect. The possibility of traveling to another universe is, however, only theoretical [33, 48].

Concentric spheres of different radii may be interpreted as various «enclosed worlds» that is enclosed in each other and connected by the common OZ axis (wormhole) (see Fig. 21, 22). At that due to Remark 3, the characteristic times (periods) increase when passing to a sphere with a greater radius and decrease when passing to a sphere with a smaller radius (see Fig. 21). Such time scale is typical for micro- and macrosystems. For instance, at the level of atoms, the period of the electron rotation about the nucleus is $\sim 10^{-8} s$, and the period of the Earth rotation about the Sun is one year.

**Conclusion**

The main result of the paper is the development of a mathematical model called by convention $\Psi$ - model, which allows describing a certain class of micro- and macrosystems.

$\Psi$ - model is based on continuum mechanics and quantum mechanics and is described on the one hand by equation of continuity (i.3), and on the other hand – by the Schrödinger equation (i.1). Thus, the solutions of one equation are the solutions of the other one and vice versa [27].

$\Psi$ - model describes micro- and macrosystems for which the vector field of probability flow velocities, charge and mass is as follows (i.9). The velocity (i.9) in a particular case describes the principle of Bohr–Sommerfeld quantization [32]. The velocity field (i.9) has spiral structure on the concentrically spherical surfaces (see Figs. 3-5). At the sphere poles and on the axis, the velocity field is not defined and has a peculiarity, and at infinite, it tends to zero.

The behavior of the $\Psi$ - model in a general case is described by unsteady singular solution (1.1.39) of the Schrödinger equation having periodicity with respect to one argument and probable periodicity with respect to the other argument (see Remark 3). The period depends on the sphere radius and tends to zero at $r \to 0$ and tends to infinite at $r \to \infty$. The density may be singular in character on the sphere axis (at the poles of concentric spheres).

The application of $\Psi$ - model to the description of elementary particles ($e, \mu, \tau$) gives naturally intrinsic magnetic moment (spin) with no need for the g-factor modification. The spin is interpreted as rotation without violating the principles of the relativity theory. One of the possible «forms» of the electron in $\Psi$ - model is a rotating toroidal ring (Paragraph 1.2), as the form cannot be isotropic (sphere) due to the presence of the preferred direction (spin). The electron «dimension» may be estimate by corresponding Compton wave length $\lambda_C^e$ in $\Psi$ - model.

The magnetic field produced by the $\Psi$ -model has only one force line (1.1.11) directed along the rotation axis and formally caused by the magnetic charge. At that, magnetic field (1.1.11) satisfies classic Maxwell's equations (see Remark 1).

The application of the $\Psi$ -model to astrophysics formally gives the interpretation of the sphere poles as «black» and «white» holes as at the poles the density tends to infinite, according to solution (1.1.31). The axis connecting the poles («black» and «white» holes) formally may be interpreted as a wormhole. The spiral structure of the velocity field in the neighborhood of the poles (see Fig. 15) is interpreted formally as a spiral galaxy (anti-galaxy). The entropy $S$ in $\Psi$-model grows in one semisphere (the neighborhood of the «white» hole) and decreases in the opposite semisphere (the neighborhood of the «black» hole); at the equator it is constant. The average entropy $H = \langle S \rangle$ is constant through the entire space (galaxy and anti-galaxy). Concentric spherical layers may be formally interpreted as «enclosed worlds». Enclosed worlds are connected by one common axis (wormhole). Due to the periodicity of solution (1.1.31), periods in each enclosed world are different and depend on a sphere radius (see Remark 3). At the micro level, periods are small and at the macro level they are big.

In conclusion, we would like to quote R. Feynman: «Physicists cannot make a conversation to any other language. If you want to learn about nature, to appreciate nature, it's necessary to understand the language that she speaks in. She offers her information only in one form; we are not so unhumble as to demand that she change before we pay any attention…
This shows again that mathematics is a deep way of expressing nature».

R. Feynman. Lecture course «The Character of Physical Law» (Lecture 2).